# Title: Nanopore sequencing of the phi X 174 genome

Authors: Andrew H. Laszlo[1], Ian M. Derrington[1], Brian C. Ross[1], Henry Brinkerhoff[1], Andrew Adey[2], Ian C. Nova[1], Jonathan M. Craig[1], Kyle W. Langford[1], Jenny Mae Samson[1], Riza Daza[2], Kenji Doering[1], Jay Shendure[2], Jens H. Gundlach[1,†]

Affiliations: [1]Department of Physics, University of Washington, Seattle, WA 98195, USA
[2]Department of Genome Sciences, University of Washington, Seattle, WA 98195, USA
[†]Corresponding author. E-mail: gundlach@uw.edu

**Abstract**: Nanopore sequencing of DNA is a single-molecule technique that may achieve long reads, low cost, and high speed with minimal sample preparation and instrumentation. Here, we build on recent progress with respect to nanopore resolution and DNA control to interpret the procession of ion current levels observed during the translocation of DNA through the pore MspA. As approximately four nucleotides affect the ion current of each level, we measured the ion current corresponding to all 256 four-nucleotide combinations (quadromers). This quadromer map is highly predictive of ion current levels of previously unmeasured sequences derived from the bacteriophage phi X 174 genome. Furthermore, we show nanopore sequencing reads of phi X 174 up to 4,500 bases in length that can be unambiguously aligned to the phi X 174 reference genome, and demonstrate proof-of-concept utility with respect to hybrid genome assembly and polymorphism detection. All methods and data are made fully available.

**Main Text:** DNA sequencing is revolutionizing biomedical and other life sciences research through its expanding scope (*1*) and has a rapidly growing presence in clinical medicine (*2*). These developments are driven in part by the successful completion of the Human Genome Project (*3*) and in part by the introduction of new sequencing technologies that have dramatically reduced the cost of DNA sequencing (*4*). Although such 'next-generation' sequencing technologies have matured considerably since early proof-of-concepts (*5-7*), nearly all remain limited to short sequence reads (with the exception of real-time sequencing from elongating polymerases (*8*)) and rely on complex, expensive instrumentation. Most platforms are also limited with respect to speed and require extensive sample preparation steps prior to sequencing.

Nanopore sequencing, independently proposed by Church and Deamer in the mid-1990s, has tremendous potential to overcome these limitations and achieve long reads, low cost, and high speed while requiring minimal sample preparation and instrumentation (*i.e.* 'tricorder'-like DNA sequencing devices) (*9-12*). However, this promise has faced substantial technical challenges, such that despite nearly 20 years of effort, nanopore-derived sequence reads that align to complex, natural DNA sequences have yet to be demonstrated.

In nanopore devices directed at DNA sequencing, a salt solution is divided into *cis* and *trans* wells by a thin membrane. A single nanometer-scale pore in the membrane connects the *cis* and *trans* wells electrically. When a voltage is applied across this membrane, ion current flows through the pore; this current provides the primary signal. DNA is negatively charged and is electrophoretically attracted into the pore. When single-stranded (ss) DNA enters the pore, it blocks some fraction of the ion current. The fraction of the ion current that is blocked depends on the identity of nucleotides within the pore (*13-15*). Key challenges of this technique are single-nucleotide resolution and control of the DNA translocation. Single-nucleotide resolution was recently enabled through the development of MspA, a protein pore with a short and narrow constriction (*11, 13, 15, 16*). DNA translocation control was also recently enabled through the use of molecular motors such as phi29 DNA Polymerase (DNAP) (*11, 17*) (**Fig. 1**).

We have found that the currents in the MspA pore are determined by about four nucleotides at any given time (*11, 15*). Each four-nucleotide combination (*i.e.* quadromer) has its own unique current value and in a few cases, nucleotides outside of a quadromer can have a small additional influence on the current. This prompted us to measure the ion current associated with each of the 256 possible quadromers.

We constructed a 256 nucleotide-long cyclical de Bruijn sequence (*18*) containing all possible combinations of four nucleotides (**Table S1**). We divided the de Bruijn sequence into eight separate strands (*19*) and synthesized these with appropriate modifications to facilitate insertion into the pore, to initiate proper polymerase function, and to allow for ion current calibration (*11, 17, 20*) (**Fig. 1a**). Phi29 DNAP-based control of translocation results in two reads from each DNA molecule: 1) 'unzipping' wherein one strand of the DNA moves 5' to 3' through the pore as the polymerase is forced to unzip the complimentary strand, and 2) 'synthesis' wherein the DNA moves 3' to 5' after the primer enters the DNAP's active site and the DNAP begins synthesizing a second complimentary strand (*11*). As the polymerase moves along the strand, one nucleotide at a time, the identity of the quadromer within the MspA pore shifts in lock-step (**Fig 1b**), resulting in discrete changes in the measured ion current (**Fig. 1c**). We performed nanopore sequencing of all eight strands, averaging signals observed across multiple molecules of each strand to estimate the current level of the 256 quadromers, *i.e.* a 'quadromer map' (**Fig. 2a, Supplemental Fig. 2-4, Supplemental Table 2**).

We next sought to evaluate whether the quadromer map constructed by nanopore sequencing of the de Bruijn sequence was predictive of current levels for previously unmeasured, natural DNA sequences. To assess this, we constructed and nanopore sequenced a genomic DNA sequencing library from the bacteriophage phi X 174. Specifically, we attached asymmetric adaptors (a nicked hairpin adaptor and a cholesterol tailed adaptor; green and orange respectively in **Fig. 1a**) to the ends of linearized phi X 174 dsDNA, and used phi29 DNAP to draw ssDNA through a mutant MspA pore in single-nucleotide steps (*11*) (**Fig. 1b,c**). The measured current level sequences were then compared to predicted current levels based on the quadromer map. **Fig. 2b** shows quadromer map-based predicted current levels versus a consensus of 22 nanopore reads for a representative ~100 nucleotide (nt) region of the phi X 174 genome.

Overall, the predicted current levels from the de Bruijn sequence-based quadromer map strongly match the observed current levels from nanopore sequencing of the phi X 174 DNA ($r = 0.9905$, 95% confidence bounds [0.9859-0.9936]). However, differences between prediction and measurement are statistically significant. Close analysis suggests that this error is dominated by shifts in the positioning of the DNA within the pore's constriction due either to DNA secondary structure within the vestibule or DNA interactions with the pore vestibule or constriction (*19*). However, as independent reads of the same sequence yield extremely reproducible current values (*11, 20*), we conclude that the small differences between prediction and observation are systematic and likely due to sequence context outside of the quadromer itself.

The strong homology between quadromer-based current predictions and nanopore sequencing reads can be used to perform alignments to reference genomes and sequence databases with high confidence. As a first assessment, we subjected three PCR amplicons derived from phi X 174 to nanopore sequencing in a blinded fashion, *i.e.* the individuals performing sequencing and analysis were not aware of the genomic positions of the amplicons. After extracting the current levels from nanopore reads using a newly developed algorithm (*19*), we aligned the observed current levels from each read to predicted current levels obtained by applying the quadromer map to the known phi X 174 genome sequence (**Fig. 3a-b**). Our alignment algorithm is similar to Needleman-Wunsch alignment (*19*, *21*, *22*) but allows for backsteps in the series of levels (**Figs. 3c,** and **Supplemental figs. 5** and **6**). We assessed confidence in these alignments by comparing alignment scores with those obtained against random sequences (**Supplemental Fig. 7** (*19*)). The vast majority (30 out of 31) of nanopore sequencing reads with a probability of false alignment below $1 \times 10^{-4}$ aligned to one of three regions; un-blinding confirmed that these corresponded to the locations along the phi X 174 genome from which the three PCR amplicons were derived (**Supplemental Figure 8**).

We next assessed whether we could achieve long nanopore sequencing reads. We constructed a genomic DNA sequencing library by ligating asymmetric adaptors to the linearized, full-length phi X 174 genome as described above, and this library was nanopore sequenced (*19*). We generated 106 long (>200 base-pair) ion current recordings corresponding to single molecules within this library. We aligned these reads to ion current levels predicted with the quadromer map; 92 of these reads aligned with high confidence to the phi X 174 genome and are shown in **Figure 4a** (misalignment probability estimated at $< 1\text{x}10^{-10}$). Within this set of aligned reads, ~60% were >1,000 bp, ~20% were >2,000 bp, and ~10% were >3,000 bp. This is in contrast to the length distribution of our library (**Supplemental Fig. 9**) which contains far longer strands implying that DNAP dissociation from the strand is the primary cause of event termination. As expected, the 5' end of most reads aligns to the cut site of the restriction enzyme used to linearize the genome, and these reads are split approximately equally between the sense and antisense strands (**Fig. 4a**). The 92 reads comprise a sum total of 118 kilobases (kb) with mean 21.9-fold coverage of the phi X 174 reference genome (range: 10-fold to 44-fold) (**Fig. 4b**).

The 10,772 bases contained within the phi X 174 sense and antisense strands include on average 35 instances of nearly all quadromers (255 out of 256) in diverse sequence contexts (**Supplemental Fig. 10**), and are likely to yield a more reliable quadromer map than the de Bruijn sequence alone. We therefore used these independent measurements of each quadromer to generate an improved quadromer map (**Supplemental Fig. 11, Supplemental Table 2**). The new quadromer map is a closer match to measured levels (**Fig. 2c**; r = 0.9936, 95% confidence bounds [0.9908-0.9958] as compared to r = 0.9905 with bounds [0.9859-0.9936]).

We then explored the potential of nanopore reads to facilitate hybrid assembly (*23-25*), by aligning short Illumina sequencing reads directly to the nanopore ion current measurements using the afore described alignment software. Specifically, we took 11,000 single-end 100 bp Illumina MiSeq shotgun reads from phi X 174 and aligned these directly to a single 3,800 bp nanopore sequencing read (**Supplemental Fig 12**). **Figure 4c** shows the alignment locations of a representative 38 Illumina reads within the nanopore read. As nanopore sequencing develops longer reads and higher throughput, such alignments may facilitate rapid and accurate sorting of short sequence reads into their proper order for *de novo* genome assembly.

To assess whether long nanopore sequencing reads could be accurately aligned against a large database of naturally occurring DNA sequences, we took one 250-level sub-region of ion currents from three individual long nanopore reads and individually aligned these 250-level regions to a 156 Mb database containing 5287 viral genomes, including phi X 174. The highest scoring alignment for all nanopore sequencing reads was to the phi X 174 genome, each with high confidence (>99.9996%, **Supplemental Fig. 13**) implying that nanopore read quality is sufficient for unambiguous species identification. These 250-level alignment 'seeds' could then be extended in both directions to the full nanopore sequencing read, yielding alignments to phi X 174 identical to the targeted alignments shown in **Fig. 4a** with high confidence.

Finally, we assessed our ability to detect single nucleotide polymorphisms (SNPs). SNPs can be detected by comparing nanopore reads to a previously measured nanopore consensus (*19, 20*) (comparison to a 'reference consensus' minimizes the impact of the systematic, context-dependent deviations from the quadromer map predictions discussed above). To systematically assess our power to detect single base substitutions, we iteratively inserted a total of 1044 'mock SNPs' to the reference genome of phi X 174, *i.e.* introducing quadromer map values corresponding to these SNPs at the appropriate locations in the phi X 174 reference consensus map(*26*). We then aligned 33 of the nanopore sequencing reads from phi X 174 to the modified reference consensus. We successfully called 77.4% of the mock SNPs (*19*). These data and methods provide a starting point for the further development of variant calling algorithms for nanopore sequencing data.

This work is the first demonstration of nanopore sequencing of long, complex, natural DNA strands. By measuring the ion current signal associated with all 256 possible 4-mers as they translocate through the constriction of the MspA nanopore, we report the first nanopore 'quadromer map.' This map is highly predictive of ion current levels of previously unmeasured, complex, natural DNA sequences. We exploit this reproducible behavior of MspA on quadromers to develop both a level finding algorithm as well as a dynamic programming alignment algorithm for nanopore sequencing reads. We apply these tools to unambiguously align long nanopore sequencing reads generated from phi X 174 to the corresponding reference genome sequence, including reads that span up to 4,500 bases in length. We then show proof-of-concept utility with respect to organismal identification as well as for hybrid *de novo* assembly. Lastly, we demonstrate algorithmic approaches for successful SNP detection using nanopore reads.

A limitation of our current system is that the amplitude of ion current levels alone does not provide enough information for direct *de novo* sequencing, *i.e.* conversion of ion currents to accurate sequences in the absence of a reference for alignment and comparison. However, additional information is contained in the variance, the duration, and the voltage dependence of each current level that may enable *de novo* sequencing with improved algorithms. Furthermore, much of the variance in current levels associated with the nanopore sequencing system described here results from the erratic and stochastic motion of the phi29 polymerase as it feeds the DNA through the pore. The fact that the ion current is affected by ~four nucleotides enables identification of polymerase skips and toggles as well as deciphering short homopolymer regions. However, switching to a different enzyme, *e.g.* a helicase, that translates along DNA monotonically and with reduced stochasticity is expected to sharply improve performance.

Importantly, the nanopore sequencing performed here was implemented on a low-cost experimental device – under $20,000 with vast potential for cost reduction with industrialization – in a small experimental lab, with essentially real-time results from a single MspA pore. Full realization of nanopore sequencing's potential will require additional progress in areas including nanopore parallelization (*27*), channel setup (*28*-*31*), and microfluidics (*32*).

Despite the remaining hurdles, our demonstration of a highly predictive quadromer map and of 4,500 bp interpretable nanopore reads – corresponding to natural DNA sequences and generated in real-time with a low-cost device – represents a major milestone in the nearly 20 year history of this technological paradigm. All experimental methods, raw data, and algorithms are made fully available to the research community to facilitate the further maturation of nanopore sequencing.

**Acknowledgments:** This work was supported by the National Institutes of Health, National Human Genome Research Institutes (NHGRI) $ 1,000 Genome Program Grants R01HG005115 and R01HG006321 (to J.H.G.) in addition to grant HG006283 from the National Genome Research Institute (to J.S.) and graduate research fellowship DGE-0718124 from the National Science Foundation (to A.A.)

## Figure 1

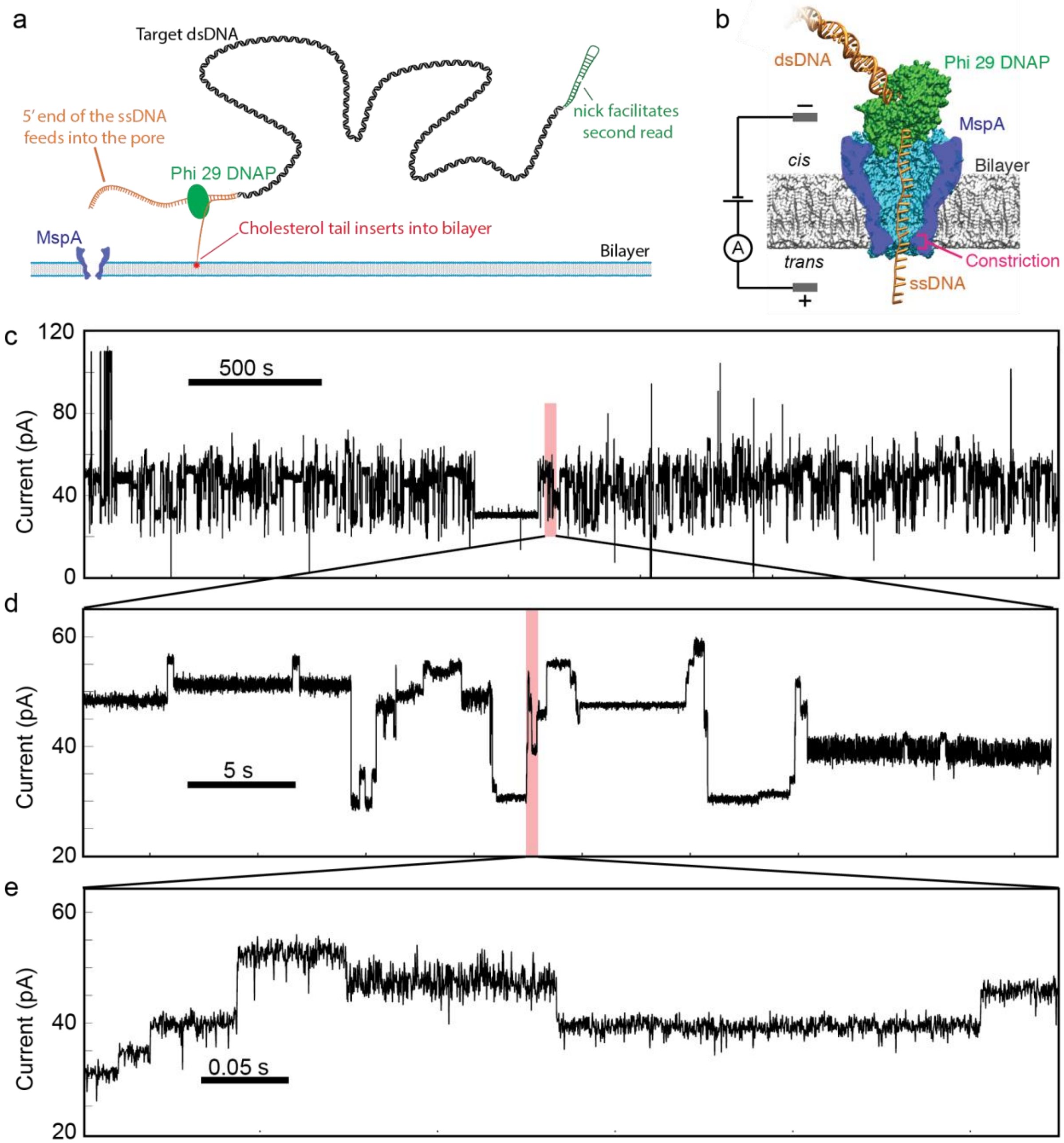

**Figure 1: Experimental schematic and raw data.** a) Method of adapting dsDNA for nanopore sequencing. The first adaptor (orange) includes a cholesterol tail which inserts into the membrane increasing DNA capture rates (*33*) while, the long 5' single stranded overhang facilitates insertion into the pore. A second adaptor (green) enables re-reading of the pore using the DNAP's synthesis mode (*11*, *17*). b) The protein nanopore MspA is shown in blue, phi 29 DNAP in green and DNA in orange. An applied voltage across the bilayer drives an ion current through the pore and an amplifier measures the current. DNA bases within the constriction determine the ion current. Phi 29 DNAP steps DNA through the pore in single-nucleotide steps. c-e) Raw data for a representative 3000-second time window. Ion current changes as DNA is fed through the pore in single-nucleotide steps. Panels d and e each show a 1% section of the preceding panel's data shaded in red.

## Figure 2

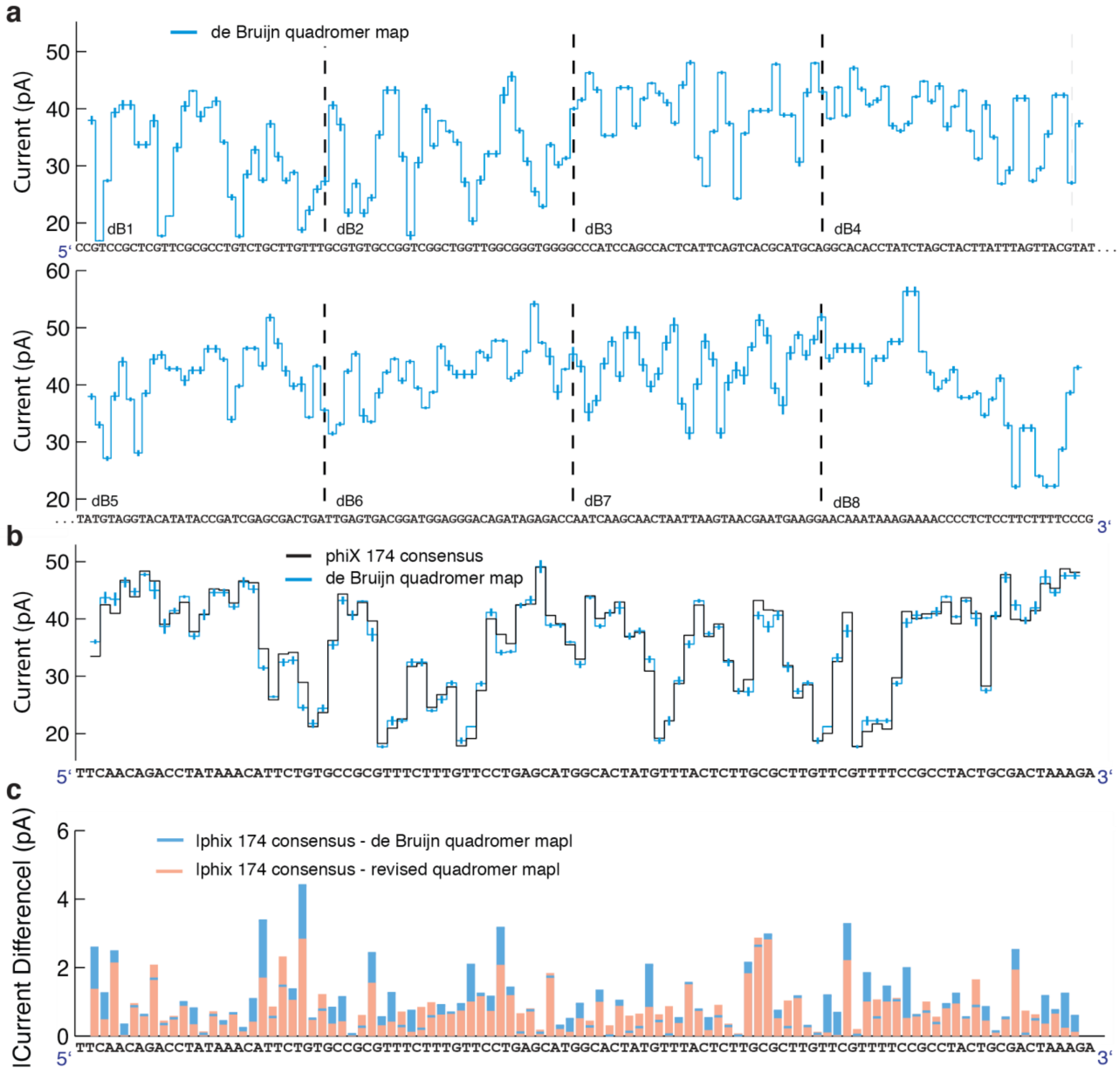

**Figure 2: A quadromer map predicts current levels for previously unmeasured DNA. a**) Current levels observed for all possible 4-nucleotide sequences (quadromers) measured in 8 segments of a 256-nucleotide de Bruijn sequence. **b**) The black trace shows a consensus based on 22 reads of phi X 174 DNA. This is compared to predicted current levels based on the de Bruijn quadromer values. Error bars are the variance of the measured quadromer values. We use a consensus to correct for insertion/deletion errors caused by the stochastic motion of the phi29 DNAP(*11*). **c**) Absolute current difference between quadromer map and measured consensus for the ~100 level sequence shown in panel **b** using the de Bruijn quadromer map (blue) and the revised quadromer map (red). In most instances, the revised map improves the predictive ability of our map. The correlation coefficient between measured values and the de Bruijn quadromer values is 0.9905 (95% confidence bounds [0.9859-0.9936]) while the correlation coefficient between measured values and the revised quadromer values is 0.9938 (95% confidence bounds [0.9908-0.9958]).

**Figure 3**

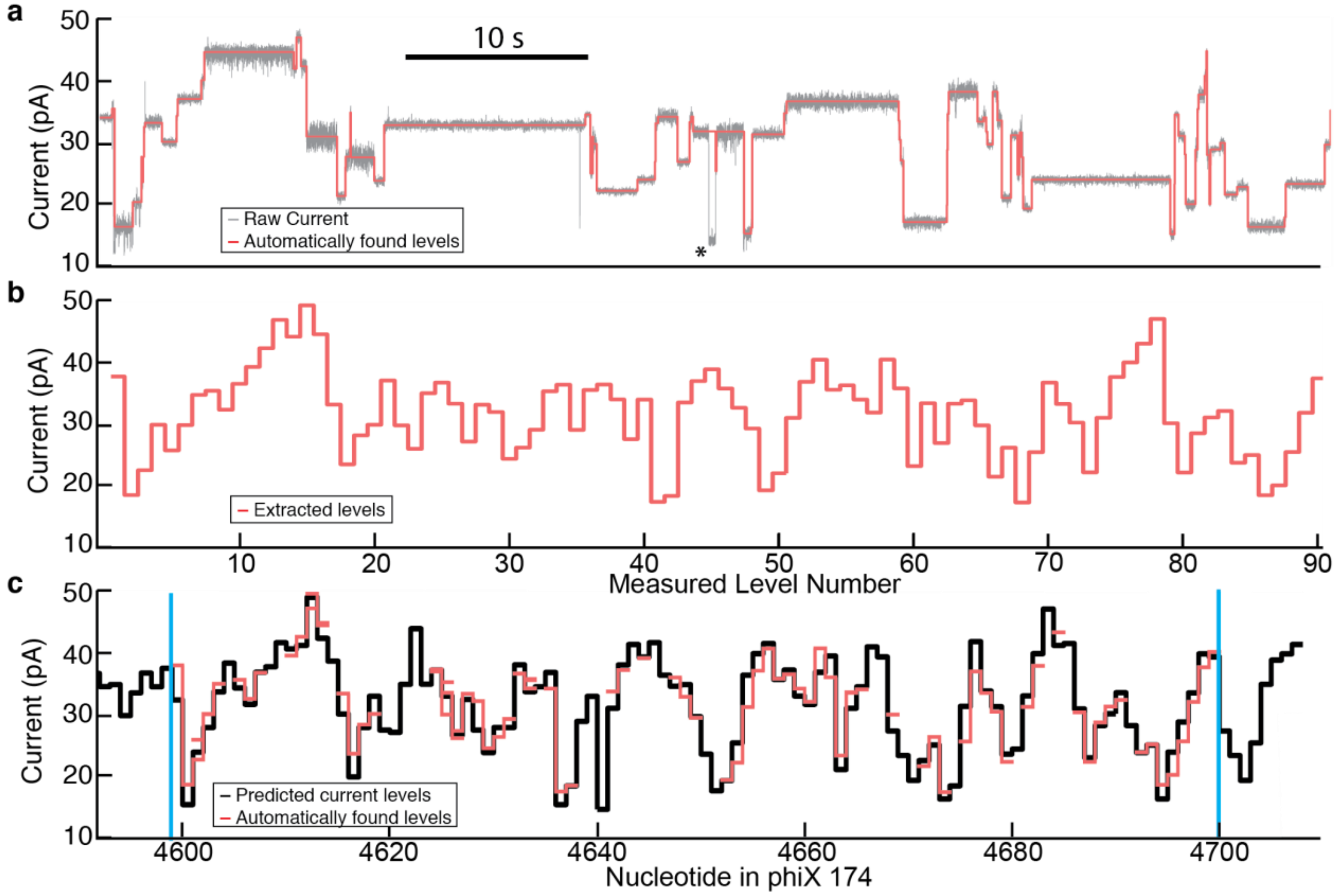

**Figure 3. Raw data to alignment a**) A level finding algorithm (*19*) is used to identify transitions between levels in the current trace. A subsequent filter removes most repeated levels due to polymerase backsteps (indicated by '*'). **b**) We extract the sequence of median current values of each level. **c**) Next, we align the current values to predicted values from the reference sequence using the quadromer map (Fig. 2A). Alignment is performed with a dynamic programming alignment algorithm similar to Needleman-Wunch alignment (*19*, *21*). In some locations, levels are skipped in the nanopore read either due to motions of the DNAP or errors made by the level finding algorithm, while in other places backsteps result in multiple reads of the same level. We determine read boundaries from the first and last matched levels in the reference sequence. Read boundaries are indicated by the blue lines. The above alignment had an estimated 6.4x $10^{-15}$ probability of false alignment.

## Figure 4

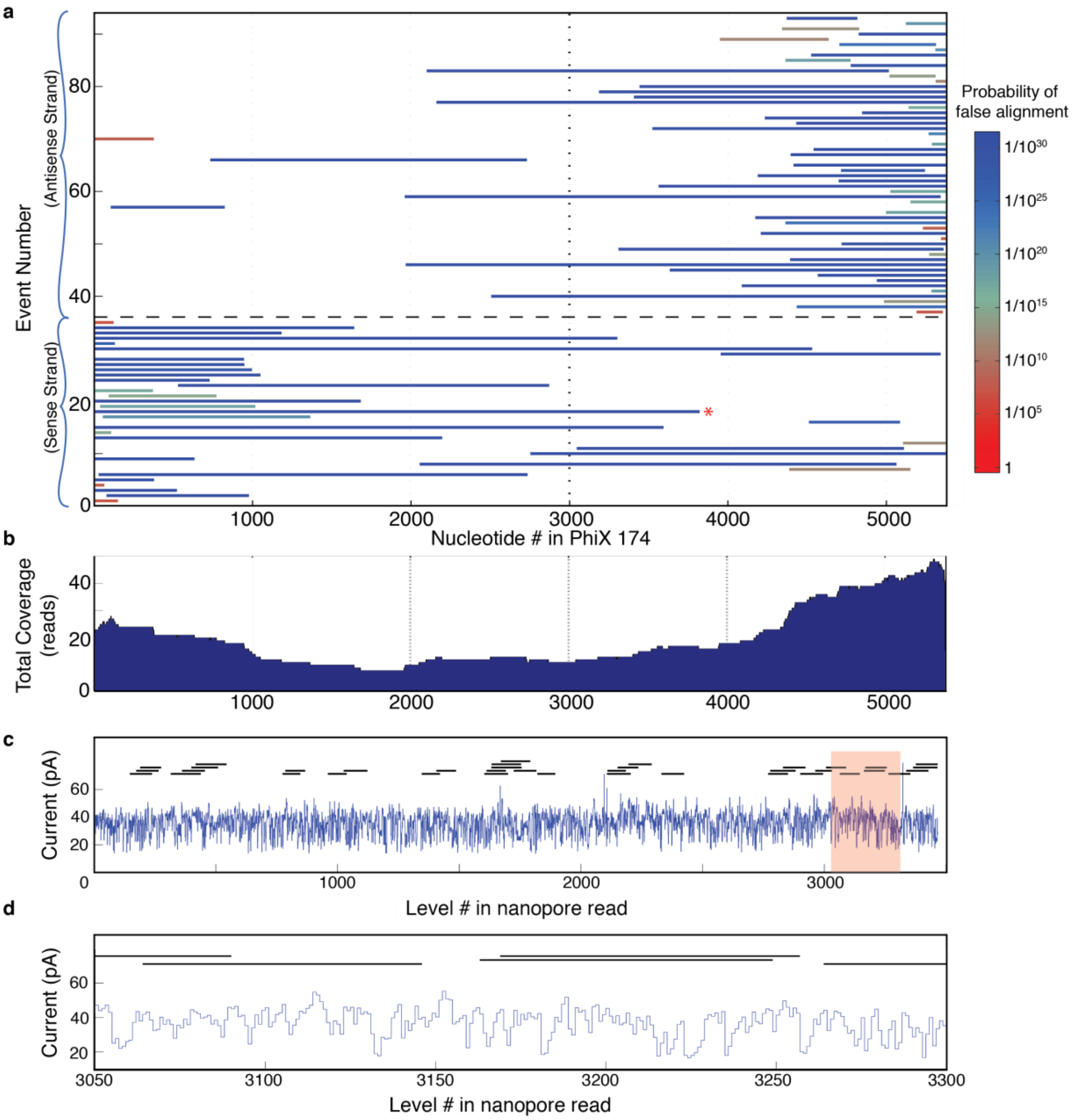

**Figure 4: Alignments to reference sequence and hybrid reconstruction** **a**) Coverage plot for 91 nanopore sequencing reads of bacteriophage phi X 174 genomic DNA. Left and right alignment bounds are indicated by the extent of the line for each read. Random attachment of the asymmetric adaptors results in reads of both sense and antisense strands. Reads below the black dashed line (events 1-38) are sense strands while reads above (events 40-92) are antisense strands. Most reads begin near the 5' end of the linearization cut site and proceed towards the 3' end as the phi29 DNAP unzips the double stranded DNA. **b**) Sum total coverage for each region within the phi X 174 genome. This graph indicates the number of reads that cover any given section of the genome using the sense and antisense strands. **c**) Hybrid assembly of Illumina sequencing reads using a single nanopore read (*19*). 38 Illumina reads (horizontal black lines) are aligned to a single 3819 nt long nanopore read (blue trace; indicated by the * in panel **a**). **d**) Detail of shaded region in panel **c**. Six 100 bp Illumina reads are shown where they align to the nanopore read. Such hybrid assembly could facilitate rapid *de novo* genome assembly of short, high quality reads requiring far lower coverage.

## Supplement for: Nanopore sequencing of the phi X 174 genome

Andrew H. Laszlo[1], Ian M. Derrington[1], Brian C. Ross[1], Henry Brinkerhoff[1], Andrew Adey[2], Ian C. Nova[1], Jonathan M. Craig[1], Kyle W. Langford[1], Jenny Mae Samson[1], Riza Daza[2], Kenji Doering[1], Jay Shendure[2], Jens H. Gundlach[1,†]

[1]Department of Physics, University of Washington, Seattle, WA 98195, USA
[2]Department of Genome Sciences, University of Washington, Seattle, WA 98195, USA
[†]Corresponding author. E-mail: gundlach@uw.edu

# Table of Contents:

**Methods and materials:**

**Pore establishment:**

A single MspA pore was established in a bilayer as previously described (*11*,*13*,*15*,*20*). Quadromer DNA was ordered from PAN labs at Stanford. DNA oligos were mixed and prepared as previously described (*11*,*20*).

**Data acquisition:**

Data was acquired with a sampling rate of 500 kHz on Axopatch 200B or Axopatch 1B amplifiers filtered at 100 kHz. Data was downsampled to 5 kHz by averaging every 100 datapoints. DNA interaction events were detected using a thresholding algorithm as previously described (*11*) and good events were selected automatically using characteristics such as duration, mean, and standard deviation. Levels within good events were then selected either by hand (for the initial quadromer data) or with an automated level-finding algorithm (for all other data). The level finding algorithm is described in detail below (see **Sup. Fig. 1** for data reduction flowchart).

**De Bruijn sequence design:**

Just as the circular eight letter sequence …AAABABBB... contains all eight three-letter combinations of A and B–AAA, AAB, ABA, BAB, ABB, BBB, BBA, BAA–one can construct a cyclical 256 letter sequence that contains all 256 four-letter combinations of A, C, G, and T (*18*).

The 256 nt long de Bruijn sequence was divided up into eight separate strands. This was to ensure DNA accuracy because high purity custom oligos longer than ~100 were not readily available. Each strand contained part of the quadromer map sequence but also contained an ion current calibration sequence that allowed us to correct for buffer evaporation and voltage offsets. In order to use the phi 29 DNAP control method (*11*,*17*), these strands also had a portion of sequence conjugated to a hairpin primer (see table S1 for strand construction and sequences). We made several measurements of each of the eight strands and aligned the extracted current levels to the known DNA sequence (see **sup. figs. 2-4**). Oligos were mixed together and annealed as previously described (*11*,*20*).

**Adaptor design:**

Adaptors were designed to each contain ½ of a NotI restriction endonuclease site to allow digestion of adaptor-dimers that were produced during the shotgun ligation. Both the fantail (FT-½NotI) and hairpin adaptors (HP-½NotI) were comprised of top and bottom oligos ordered from IDT (FT-½NotI-top: 5'- (Phosphate) (3 Carbon Spacer) AAA AAA ACC TTC C (3 Carbon Spacer) CCT TCC CAT CAT CAT CAG ATC TCA CGC GG -3’, FT-½NotI-bot: 5’- (Phosphate) GGC GCA CTC TAG ACT TTT TAA ATT TGG GTT T (3 Carbon Spacer) (Cholesterol) -3’, HP-½NotI-top: 5’- (Phosphate) CGC CTA CGG TTT TTC CGT AGG CGT ACG C (Uracil) TAC TTG TAC TTG GCG G -3’, HP-½NotI-bot: 5’- (Phosphate) CCG CCA AGT ACA AGT AAG CGT A -3’). The cholesterol tag at the end of the fantail adaptor causes the DNA to bind to the bilayer thereby increasing the DNA concentration near the pore and increasing DNA-pore interactions (*33*).

Oligos were resuspended to 100 µM in 10 mM Tris, and annealed by combining 20 µL of the top and bottom oligos with the addition of 60 µL 10 mM Tris followed by heating to 95°C for 2 minutes and gradual cooling to room temperature in a polystyrene casing. The two-oligo scheme for the hairpin adaptor was designed to prevent synthesis of excessively long oligos. The nick present between the oligos ligates during the adaptor ligation process, followed by subsequent introduction of the desired nick by digestion of the uracil base with USER enzyme.

**Library construction:**

Phi X 174 nanopore libraries were constructed using a shotgun-ligation approach. 0.5 to 4 µg of Phi X 174 gDNA (Thermo Scientific) was restriction-digested using 5U of SspI (NEB) in 1× SspI Reaction Buffer for 2 hours at 37°C to linearize DNA and produce blunt ends followed by SPRI bead purification. DNA was resuspended in 42 µL Elution Buffer (EB, QIAGEN) followed by addition of 5 µL 10× T4 DNA Ligase Buffer (NEB), 1 µL of each annealed adaptor (FT-½NotI and HP-½NotI), and 1 µL (400U) of T4 DNA Ligase (NEB) and incubated overnight at 16°C. Ligase was heat-inactivated at 65°C for 15 minutes followed by cooling on ice and addition of 1 µL 10× T4 DNA Ligase Buffer, 1 µL (1U) USER Enzyme (NEB), 1 µL (20U) NotI-HF (NEB), and 2 µL (10U) λ Exonuclease (NEB) and incubation at 37°C for 2 hours. DNA was then purified using 45 µL SPRI beads. A subset of samples were gel-size-selected to remove adaptor-dimer bands on a 1% agarose gel (SeaKem) and purified using the column-based Gel Purification Kit (QIAGEN) and eluted in EB.

**Alignment Algorithm:**

Alignments were performed using a novel dynamic programming algorithm described later in the supporting text. Quality scores for alignments were estimated by comparing the maximal alignment score to the alignment scores obtained from alignments of measured strands to random DNA sequences with the same GC content as that of phi X 174. A data processing flowchart is available below (**Sup. Fig. 1**).

## Supplemental Figure 1: Data reduction flow chart

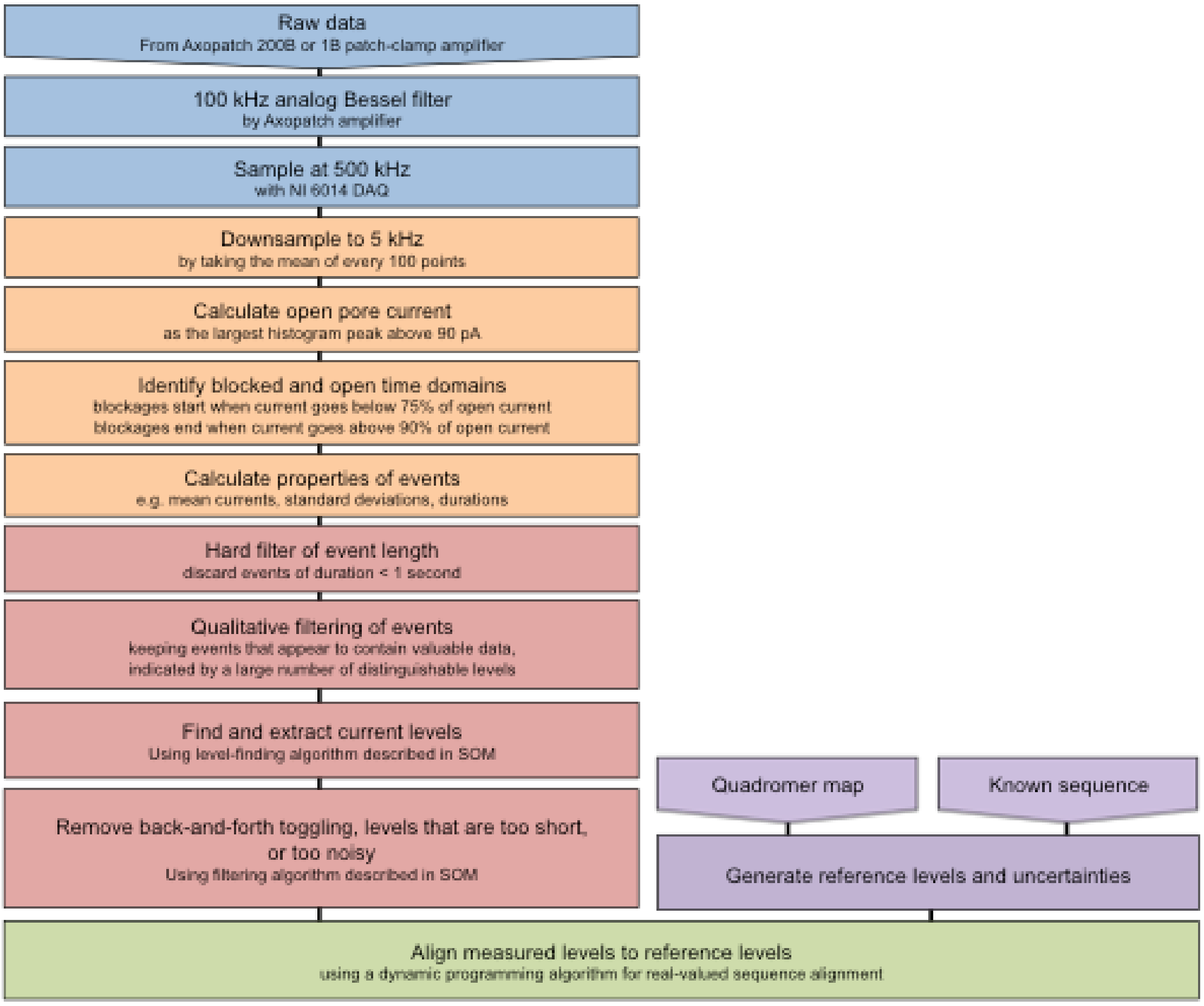

**Supplemental Table 1.1: DNA strands used in this study:**

| Construct | Sequence | # of Pores | # of Events |
|---|---|---|---|
| **deBruijn Section 1** | 5' PAAAAAAACCTTCCXTTTTCCCGTCCGCTCGTTCGCGCCTGTCTGCTTGTTTGCGTGTGCCGGTCGGCTAAGCATTCTCATGCAGGTCGTAGCC 3' | 4 | 7 |
| Nucleotide number | 0 5 10 15 20 25 30 35 40 45 50 55 60 65 70 75 80 85 90 | | |
| DNA purpose | (--Calibration--)(-----------------------Sequence------------------)(--Blocker--)(--Primer--) | | |
| **deBruijn Section 1 Blocker** | 5' TTAGCCGACCGGCACACGCAAACAAGCAGACAXXXXXXZ 3' | | |
| Nucleotide number | 0 5 10 15 20 25 30 35 40 | | |
| | | | |
| **deBruijn Section 2** | 5' PAAAAAAACCTTCCXTGTTTGCGTGTGCCGGTCGGCTGGTTGGCGGGTGGGCCCATCAAAACACTCATAAGCATTCTCATGCAGGTCGTAGCC 3' | 7 | 14 |
| Nucleotide number | 0 5 10 15 20 25 30 35 40 45 50 55 60 65 70 75 80 85 90 | | |
| DNA purpose | (--Calibration--)(-----------------------Sequence------------------)(--Blocker--)(--Primer--) | | |
| **deBruijn Section 2 Blocker** | 5' TTATGAGTGTTTTGATGGGCCCACCCGCCAAXXXXXXZ 3' | | |
| Nucleotide number | 0 5 10 15 20 25 30 35 40 | | |
| | | | |
| **deBruijn Section 3** | 5' PAAAAAAACCTTCCXGTGGGCCCATCCAGCCACTCATTCAGTCACGCATGCAGGCACACCTATCTAGCTAAGCATTCTCATGCAGGTCGTAGCC 3' | 5 | 12 |
| Nucleotide number | 0 5 10 15 20 25 30 35 40 45 50 55 60 65 70 75 80 85 90 | | |
| DNA purpose | (--Calibration--)(-----------------------Sequence------------------)(--Blocker--)(--Primer--) | | |
| **deBruijn Section 3 Blocker** | 5' TTAGCTAGATAGGTGTGCCTGCATGCGTGACTXXXXXXZ 3' | | |
| Nucleotide number | 0 5 10 15 20 25 30 35 40 | | |
| | | | |
| **deBruijn Section 4** | 5' PAAAAAAACCTTCCXATGCAGGCACACCTATCTAGCTACTTATTTAGTTACGTATGTAGGTACATATACAAGCATTCTCATGCAGGTCGTAGCC 3' | 5 | 11 |
| Nucleotide number | 0 5 10 15 20 25 30 35 40 45 50 55 60 65 70 75 80 85 90 | | |
| DNA purpose | (--Calibration--)(-----------------------Sequence------------------)(--Blocker--)(--Primer--) | | |
| **deBruijn Section 4 Blocker** | 5' TTGTATATGTACCTACATACGTAACTAAATAAXXXXXXXZ 3' | | |
| Nucleotide number | 0 5 10 15 20 25 30 35 40 | | |
| | | | |
| **Legend** | | | |
| X = abasic site | | | |
| Z = 3-carbon spacer | | | |
| P = phosphate group | | | |

## Supplemental Table 1.2: DNA strands used in this study:

| Construct | Sequence | # of Pores | # of Events |
|---|---|---|---|
| **deBruijn Section 5** | 5' PAAAAAAACCTTCCXTTACGTATGTAGGTACATATACCGATCGAGCGACTGATTGAGTGACGGATGGAGAAGCATTCTCATGCAGGTCGTAGCC 3' | 6 | 12 |
| Nucleotide number | 0 5 10 15 20 25 30 35 40 45 50 55 60 65 70 75 80 85 90 | | |
| DNA purpose | (--Calibration--)(----------------------Sequence------------------)(--Blocker--)(--Primer--) | | |
| **deBruijn Section 5 Blocker** | 5' TTCTCCATCCGTCACTCAATCAGTCGCTCGATXXXXXXXZ 3' | | |
| Nucleotide number | 0 5 10 15 20 25 30 35 40 | | |
| | | | |
| **deBruijn Section 6** | 5' PAAAAAAACCTTCCXACTGATTGAGTGACGGATGGAGGGACAGATAGAGACCAATCAAGCAACTAATAAAAGCATTCTCATGCAGGTCGTAGCC 3' | 3 | 15 |
| Nucleotide number | 0 5 10 15 20 25 30 35 40 45 50 55 60 65 70 75 80 85 90 | | |
| DNA purpose | (--Calibration--)(----------------------Sequence------------------)(--Blocker--)(--Primer--) | | |
| **deBruijn Section 6 Blocker** | 5' TTTTATTAGTTGCTTGATTXXXXXXZ 3' | | |
| Nucleotide number | 0 5 10 15 20 25 | | |
| | | | |
| **deBruijn Section 7** | 5' PAAAAAAACCTTCCXAGACCAATCAAGCAACTAATTAAGTAACGAATGAAGGAACAAATAAAGAAAACCAAGCATTCTCATGCAGGTCGTAGCC 3' | 8 | 11 |
| Nucleotide number | 0 5 10 15 20 25 30 35 40 45 50 55 60 65 70 75 80 85 90 | | |
| DNA purpose | (--Calibration--)(----------------------Sequence------------------)(--Blocker--)(--Primer--) | | |
| **deBruijn Section 7 Blocker** | 5' TTGGTTTTCTTTATTTGTTCCTTCATTCGTTACTTAXXXXXXXZ 3' | | |
| Nucleotide number | 0 5 10 15 20 25 30 35 40 45 | | |
| | | | |
| **deBruijn Section 8** | 5' PAAAAAAACCTTCCXGAAGGAACAAATAAAGAAAACCCCTCTCCTTCTTTTCCCGTCCGCTCGTTCGCGAAGCATTCTCATGCAGGTCGTAGCC 3' | 1 | 17 |
| Nucleotide number | 0 5 10 15 20 25 30 35 40 45 50 55 60 65 70 75 80 85 90 | | |
| DNA purpose | (--Calibration--)(----------------------Sequence------------------)(--Blocker--)(--Primer--) | | |
| **deBruijn Section 8 Blocker** | 5' TTCGCGAACGAGCGGACGGXXXXXXXZ 3' | | |
| Nucleotide number | 0 5 10 15 20 25 | | |
| | | | |
| **deBruijn Primer** | 5' GCGTACGCCTACGGTTTTCCGTAGGCGTACGCGGCTACGACCTGCATGAGAATGC 3' | | |
| Nucleotide number | 0 5 10 15 20 25 30 35 40 45 50 55 | | |
| | | | |
| **Phi X Adaptor** | 5' PZAAAAAAACCTTCCZCCTTCCCATCATCATCAGATCTCACGCGG 3' | | |
| Nucleotide number | 0 5 10 15 20 25 30 35 40 45 | | |
| | | | |
| **Phi X Hairpin (1)** | 5' PCGCCTACGGTTTTTCCGTAGGCGTACGCUTACTTGTACTTGGCGG 3' | | |
| Nucleotide number | 0 5 10 15 20 25 30 35 40 45 | | |
| | | | |
| **Phi X Hairpin (2)** | 5' CCGCCAAGTACAAGTAAGCGTA 3' | | |
| Nucleotide number | 0 5 10 15 20 25 | | |
| **Legend** | | | |
| X = abasic site | | | |
| Z = 3-carbon spacer | | | |
| P = phosphate group | | | |

## Supplemental Figure 2: Current consensuses for all de Bruijn strands 1-3

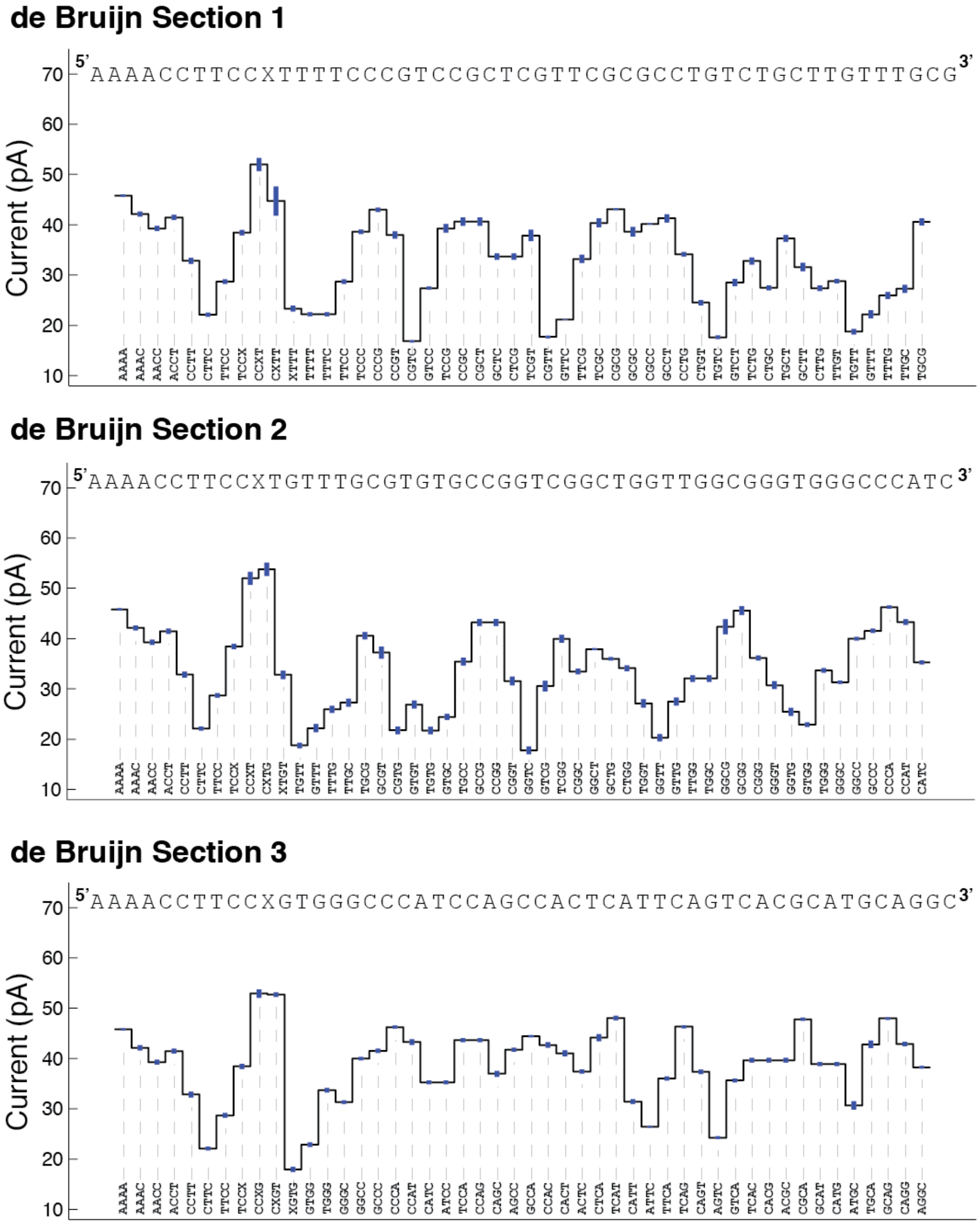

**Supplemental figure 2**: **de Bruijn sequence segments 1-3:** Consensus current level sequences and associated quadromers for de Bruijn segments 1-3. Consensuses for each strand were generated from 17, 11, and 15 reads of strands 1,2, and 3, respectively.

**Supplemental Figure 3: Current consensuses for all de Bruijn strands 4-6**

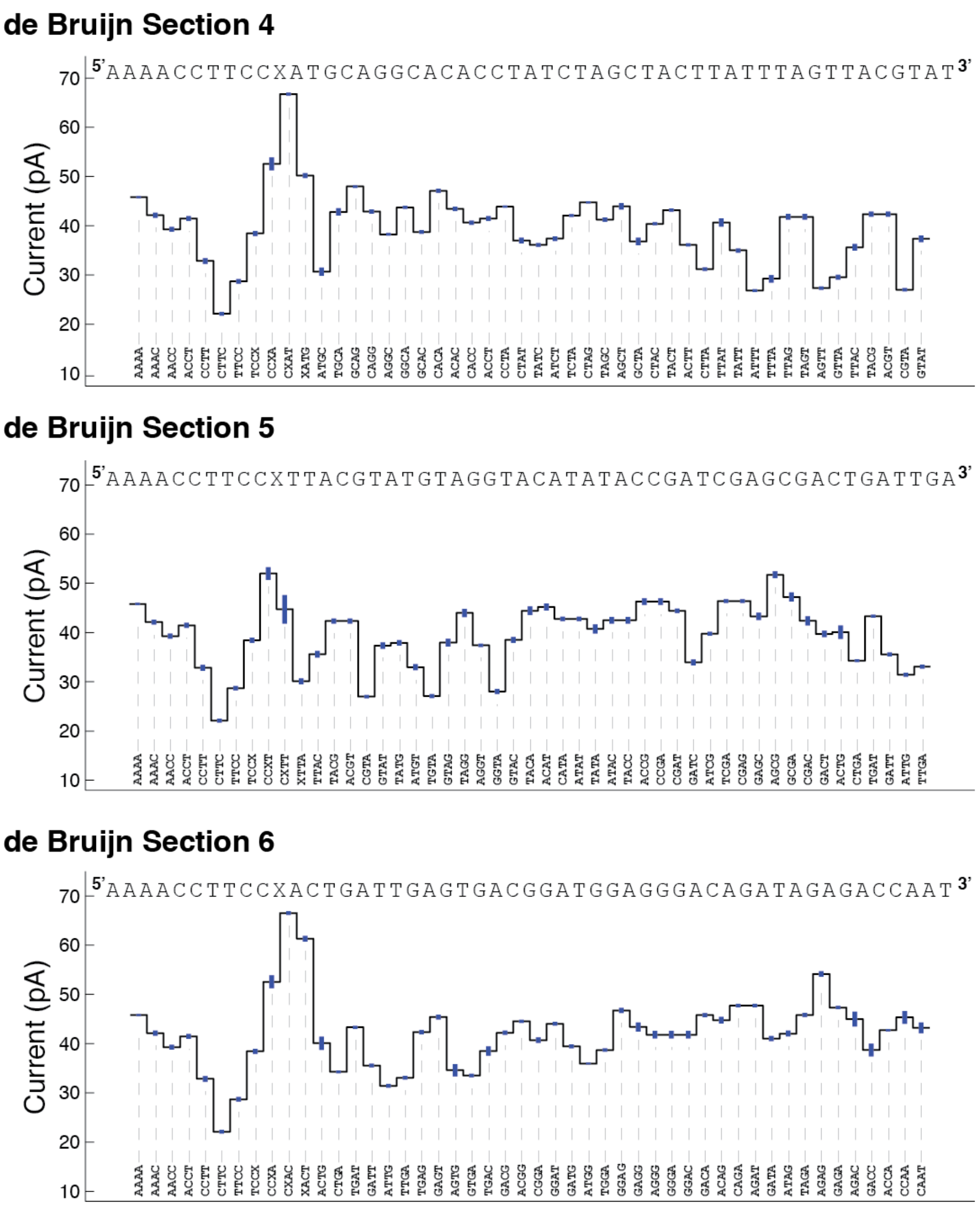

**Supplemental figure 3**: **de Bruijn segments 4-6:**Consensus current level sequences and associated quadromers for de Bruijn segments 4-6. Consensuses for each strand were generated from 12, 11, and 12 reads of strands 4,5, and 6, respectively.

## Supplemental Figure 4: Current consensuses for all de Bruijn strands 7-8

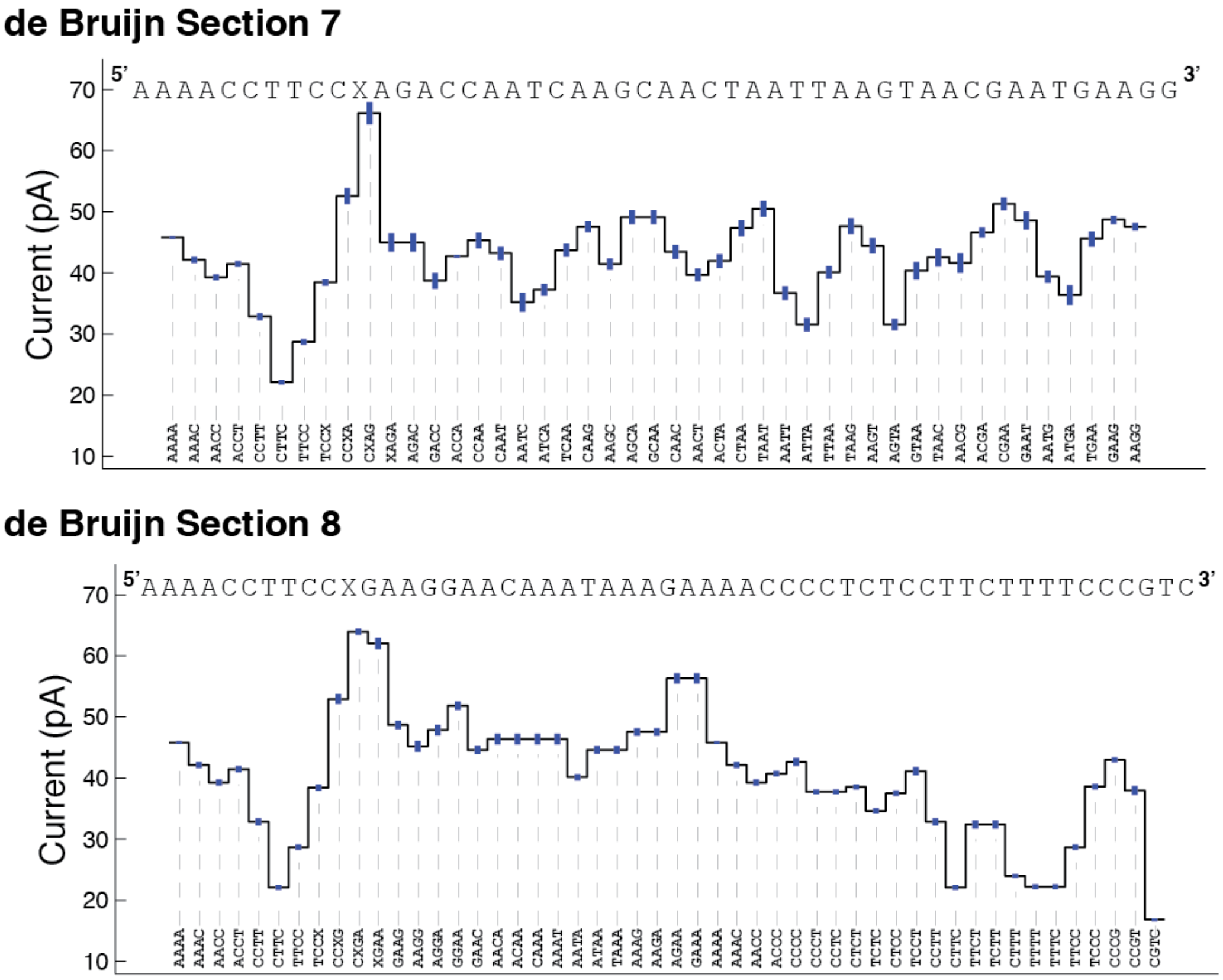

**Supplemental figure 4**: **de Bruijn segments 7-8.** Consensus current level sequences and associated quadromers for de Bruijn segments 7-8. Consensuses for each strand were generated from 14 and 7 reads of strands 7 and 8, respectively.

**Supplemental table 2.1: Table of quadromer map values beginning with A.**

| Quadromer | de Bruijn Value(pA) | Revised Value(pA) | Error on Revised Value (pA) | Number of Phi X 174 Measurements |
|---|---|---|---|---|
| AAAA | 45.8 | 46.2 | 1.4 | 65 |
| AAAC | 42.1 | 42.1 | 1.3 | 45 |
| AAAG | 47.5 | 48.5 | 1.4 | 34 |
| AAAT | 46.4 | 47.3 | 1.7 | 52 |
| AACA | 46.4 | 46.8 | 1.1 | 27 |
| AACC | 39.3 | 38.7 | 0.8 | 45 |
| AACG | 41.6 | 41.4 | 0.9 | 42 |
| AACT | 39.7 | 39.7 | 1.0 | 57 |
| AAGA | 47.5 | 48.0 | 1.2 | 42 |
| AAGC | 41.5 | 41.4 | 1.2 | 25 |
| AAGG | 45.2 | 45.2 | 1.2 | 25 |
| AAGT | 44.4 | 43.0 | 1.6 | 36 |
| AATA | 40.2 | 40.1 | 0.9 | 44 |
| AATC | 35.2 | 34.1 | 1.5 | 34 |
| AATG | 39.4 | 38.7 | 1.4 | 29 |
| AATT | 36.7 | 35.4 | 1.8 | 54 |
| ACAA | 46.4 | 46.4 | 1.3 | 34 |
| ACAC | 43.4 | 43.1 | 0.8 | 15 |
| ACAG | 44.7 | 44.8 | 1.0 | 23 |
| ACAT | 45.2 | 45.9 | 1.4 | 16 |
| ACCA | 42.7 | 43.0 | 1.0 | 50 |
| ACCC | 40.7 | 41.0 | 0.9 | 14 |
| ACCG | 46.3 | 45.4 | 0.7 | 40 |
| ACCT | 41.5 | 41.5 | 1.0 | 34 |
| ACGA | 46.6 | 46.5 | 0.6 | 48 |
| ACGC | 39.7 | 39.9 | 1.3 | 31 |
| ACGG | 44.5 | 44.4 | 0.8 | 26 |
| ACGT | 42.3 | 41.6 | 1.5 | 30 |
| ACTA | 41.9 | 42.1 | 1.3 | 43 |
| ACTC | 37.4 | 37.1 | 0.8 | 41 |
| ACTG | 40.1 | 39.3 | 1.4 | 42 |
| ACTT | 36.1 | 35.3 | 1.7 | 48 |
| AGAA | 56.3 | 55.1 | 1.6 | 43 |
| AGAC | 45.0 | 44.6 | 1.2 | 36 |
| AGAG | 54.1 | 53.0 | 1.3 | 29 |
| AGAT | 47.7 | 49.3 | 2.2 | 19 |
| AGCA | 49.1 | 48.9 | 1.5 | 26 |
| AGCC | 41.8 | 41.0 | 1.4 | 26 |
| AGCG | 51.7 | 50.4 | 1.4 | 26 |
| AGCT | 44.0 | 43.1 | 1.1 | 20 |
| AGGA | 47.9 | 47.7 | 1.1 | 26 |
| AGGC | 38.3 | 38.8 | 1.4 | 13 |
| AGGG | 41.8 | 40.6 | 2.0 | 11 |
| AGGT | 37.4 | 37.6 | 1.3 | 39 |
| AGTA | 31.5 | 31.4 | 1.4 | 53 |
| AGTC | 24.3 | 23.3 | 1.8 | 50 |
| AGTG | 34.6 | 33.8 | 1.9 | 30 |
| AGTT | 27.3 | 25.2 | 1.4 | 60 |
| ATAA | 44.6 | 44.5 | 0.9 | 37 |
| ATAC | 42.5 | 41.5 | 1.0 | 33 |
| ATAG | 42.0 | 42.8 | 0.8 | 41 |
| ATAT | 42.8 | 42.3 | 0.9 | 16 |
| ATCA | 37.2 | 37.5 | 1.1 | 20 |
| ATCC | 35.3 | 34.4 | 1.1 | 17 |
| ATCG | 39.8 | 39.3 | 1.0 | 28 |
| ATCT | 37.4 | 37.3 | 0.9 | 26 |
| ATGA | 36.4 | 36.4 | 1.1 | 25 |
| ATGC | 30.7 | 31.0 | 1.0 | 25 |
| ATGG | 36.0 | 35.6 | 0.8 | 29 |
| ATGT | 33.0 | 31.7 | 1.4 | 14 |
| ATTA | 31.6 | 31.1 | 0.9 | 46 |
| ATTC | 26.4 | 25.0 | 1.0 | 33 |
| ATTG | 31.4 | 30.6 | 1.2 | 39 |

**Supplemental table 2.2: Table of quadromer map values beginning with C.**

| Quadromer | de Bruijn Value(pA) | Revised Value(pA) | Error on Revised Value (pA) | Number of Phi X 174 Measurements |
|---|---|---|---|---|
| CAAA | 46.4 | 46.1 | 1.1 | 36 |
| CAAC | 43.5 | 43.1 | 1.4 | 41 |
| CAAG | 47.5 | 46.7 | 1.3 | 30 |
| CAAT | 43.2 | 43.3 | 1.2 | 31 |
| CACA | 47.1 | 46.9 | 1.3 | 15 |
| CACC | 40.6 | 40.3 | 1.1 | 20 |
| CACG | 39.7 | 39.4 | 1.0 | 18 |
| CACT | 41.1 | 40.8 | 0.8 | 20 |
| CAGA | 47.7 | 47.8 | 1.5 | 22 |
| CAGC | 37.0 | 37.1 | 1.2 | 31 |
| CAGG | 42.9 | 43.0 | 0.8 | 18 |
| CAGT | 37.4 | 39.1 | 1.7 | 51 |
| CATA | 42.8 | 42.6 | 1.0 | 31 |
| CATC | 35.3 | 35.1 | 1.3 | 25 |
| CATG | 38.9 | 39.4 | 1.0 | 18 |
| CATT | 31.4 | 33.1 | 2.0 | 34 |
| CCAA | 45.3 | 45.1 | 1.0 | 34 |
| CCAC | 42.7 | 42.3 | 0.7 | 15 |
| CCAG | 43.7 | 44.0 | 1.1 | 29 |
| CCAT | 43.3 | 44.5 | 1.3 | 36 |
| CCCA | 46.3 | 45.5 | 1.1 | 15 |
| CCCC | 42.6 | 42.4 | 1.2 | 10 |
| CCCG | 43.0 | 43.3 | 1.2 | 7 |
| CCCT | 37.8 | 38.0 | 1.4 | 19 |
| CCGA | 46.3 | 46.2 | 0.9 | 28 |
| CCGC | 40.7 | 40.7 | 1.1 | 40 |
| CCGG | 43.2 | 44.2 | 0.8 | 19 |
| CCGT | 38.0 | 37.9 | 1.9 | 33 |
| CCTA | 43.9 | 43.8 | 0.9 | 26 |
| CCTC | 37.8 | 37.4 | 0.9 | 27 |
| CCTG | 34.1 | 35.2 | 1.7 | 18 |
| CCTT | 32.8 | 33.6 | 1.7 | 38 |
| CGAA | 51.3 | 50.6 | 0.9 | 58 |
| CGAC | 42.4 | 41.8 | 1.2 | 44 |
| CGAG | 46.4 | 46.1 | 0.7 | 32 |
| CGAT | 44.4 | 44.1 | 1.1 | 19 |
| CGCA | 47.8 | 46.4 | 1.4 | 45 |
| CGCC | 40.2 | 39.9 | 1.0 | 31 |
| CGCG | 43.1 | 43.5 | 0.9 | 26 |
| CGCT | 40.7 | 40.9 | 0.8 | 30 |
| CGGA | 40.7 | 39.6 | 1.3 | 28 |
| CGGC | 33.5 | 33.8 | 1.0 | 25 |
| CGGG | 36.2 | 36.7 | 1.8 | 9 |
| CGGT | 31.6 | 31.8 | 2.0 | 43 |
| CGTA | 27.0 | 26.2 | 1.5 | 41 |
| CGTC | 16.8 | 16.6 | 1.6 | 49 |
| CGTG | 21.8 | 21.4 | 1.4 | 19 |
| CGTT | 17.7 | 17.9 | 1.3 | 58 |
| CTAA | 47.3 | 46.5 | 1.3 | 48 |
| CTAC | 40.4 | 40.1 | 0.9 | 34 |
| CTAG | 44.8 | 44.4 | 1.8 | 1 |
| CTAT | 37.0 | 37.5 | 1.1 | 40 |
| CTCA | 44.1 | 43.6 | 0.9 | 30 |
| CTCC | 37.5 | 37.2 | 1.0 | 27 |
| CTCG | 35.7 | 35.7 | 2.3 | 32 |
| CTCT | 38.6 | 38.2 | 0.9 | 34 |
| CTGA | 34.3 | 34.5 | 1.1 | 24 |
| CTGC | 27.5 | 27.9 | 1.0 | 38 |
| CTGG | 34.1 | 33.8 | 1.0 | 23 |
| CTGT | 24.5 | 26.1 | 2.1 | 22 |
| CTTA | 31.2 | 30.3 | 1.0 | 36 |
| CTTC | 22.1 | 22.3 | 0.9 | 52 |
| CTTG | 27.4 | 27.3 | 0.8 | 33 |
| CTTT | 24.0 | 23.6 | 1.1 | 66 |

**Supplemental table 2.3: Table of quadromer map values beginning with G.**

| Quadromer | de Bruijn Value(pA) | Revised Value(pA) | Error on Revised Value (pA) | Number of Phi X 174 Measurements |
|---|---|---|---|---|
| GAAA | 56.3 | 54.3 | 1.3 | 50 |
| GAAC | 44.6 | 44.1 | 1.1 | 44 |
| GAAG | 48.7 | 49.3 | 1.9 | 38 |
| GAAT | 48.6 | 50.0 | 1.6 | 31 |
| GACA | 45.8 | 45.1 | 1.2 | 24 |
| GACC | 38.7 | 38.6 | 0.9 | 30 |
| GACG | 42.2 | 42.4 | 0.9 | 38 |
| GACT | 39.7 | 40.4 | 1.5 | 49 |
| GAGA | 47.4 | 49.1 | 1.2 | 23 |
| GAGC | 43.3 | 43.4 | 1.2 | 21 |
| GAGG | 43.4 | 43.8 | 1.3 | 16 |
| GAGT | 45.4 | 45.2 | 1.8 | 49 |
| GATA | 41.0 | 40.5 | 0.9 | 16 |
| GATC | 33.9 | 33.9 | 2.3 | 6 |
| GATG | 39.4 | 39.2 | 1.3 | 18 |
| GATT | 35.6 | 35.1 | 1.9 | 24 |
| GCAA | 49.1 | 48.8 | 0.8 | 41 |
| GCAC | 38.7 | 39.1 | 1.5 | 22 |
| GCAG | 48.0 | 47.8 | 1.4 | 35 |
| GCAT | 38.9 | 38.8 | 1.3 | 21 |
| GCCA | 44.4 | 44.7 | 0.7 | 22 |
| GCCC | 41.5 | 41.2 | 1.4 | 8 |
| GCCG | 43.2 | 43.9 | 1.0 | 41 |
| GCCT | 41.3 | 41.5 | 0.8 | 21 |
| GCGA | 47.2 | 47.3 | 0.6 | 23 |
| GCGC | 38.6 | 38.8 | 1.2 | 22 |
| GCGG | 45.6 | 45.5 | 1.0 | 29 |
| GCGT | 37.2 | 38.1 | 1.9 | 35 |
| GCTA | 36.8 | 37.1 | 1.3 | 21 |
| GCTC | 33.7 | 34.2 | 1.4 | 20 |
| GCTG | 36.0 | 36.6 | 1.3 | 26 |
| GCTT | 31.6 | 32.9 | 1.6 | 33 |
| GGAA | 51.8 | 52.2 | 1.6 | 32 |
| GGAC | 41.8 | 41.5 | 1.6 | 26 |
| GGAG | 46.7 | 47.4 | 1.2 | 27 |
| GGAT | 44.1 | 44.3 | 1.5 | 11 |
| GGCA | 43.7 | 43.6 | 1.7 | 18 |
| GGCC | 40.0 | 39.3 | 1.6 | 8 |
| GGCG | 42.3 | 42.2 | 1.5 | 22 |
| GGCT | 37.9 | 38.3 | 1.8 | 17 |
| GGGA | 41.8 | 41.3 | 1.8 | 14 |
| GGGC | 31.3 | 31.7 | 1.5 | 6 |
| GGGG | 30.2 | 31.1 | 4.6 | 7 |
| GGGT | 30.7 | 30.9 | 2.4 | 15 |
| GGTA | 28.0 | 27.8 | 1.8 | 44 |
| GGTC | 17.8 | 17.1 | 1.8 | 40 |
| GGTG | 25.5 | 24.7 | 1.3 | 24 |
| GGTT | 20.3 | 19.9 | 1.6 | 54 |
| GTAA | 40.3 | 39.3 | 1.7 | 47 |
| GTAC | 38.5 | 37.5 | 1.9 | 31 |
| GTAG | 38.0 | 38.3 | 1.4 | 42 |
| GTAT | 37.4 | 37.2 | 1.7 | 44 |
| GTCA | 35.7 | 35.9 | 2.4 | 34 |
| GTCC | 27.4 | 26.3 | 2.1 | 31 |
| GTCG | 30.6 | 29.2 | 2.0 | 58 |
| GTCT | 28.5 | 27.1 | 1.7 | 46 |
| GTGA | 33.5 | 31.7 | 2.0 | 23 |
| GTGC | 24.4 | 24.9 | 1.8 | 24 |
| GTGG | 22.9 | 24.0 | 1.9 | 25 |
| GTGT | 26.9 | 26.9 | 1.4 | 21 |
| GTTA | 29.5 | 27.8 | 1.9 | 49 |
| GTTC | 21.2 | 20.1 | 1.8 | 47 |
| GTTG | 27.5 | 25.9 | 1.6 | 45 |
| GTTT | 22.2 | 21.4 | 1.2 | 65 |

## Supplemental table 2.4: Table of quadromer map values beginning with T.

| Quadromer | de Bruijn Value(pA) | Revised Value(pA) | Error on Revised Value (pA) | Number of Phi X 174 Measurements |
|---|---|---|---|---|
| TAAA | 44.6 | 44.7 | 1.2 | 56 |
| TAAC | 42.5 | 42.4 | 0.9 | 39 |
| TAAG | 47.6 | 47.5 | 1.0 | 28 |
| TAAT | 50.5 | 48.7 | 1.2 | 48 |
| TACA | 44.4 | 44.3 | 1.4 | 24 |
| TACC | 42.5 | 42.0 | 1.2 | 41 |
| TACG | 42.3 | 42.5 | 1.0 | 37 |
| TACT | 43.2 | 43.2 | 0.8 | 47 |
| TAGA | 45.8 | 45.3 | 1.1 | 32 |
| TAGC | 41.3 | 41.4 | 1.0 | 20 |
| TAGG | 44.0 | 44.8 | 0.7 | 29 |
| TAGT | 41.8 | 42.9 | 1.3 | 56 |
| TATA | 40.7 | 40.9 | 0.9 | 34 |
| TATC | 36.1 | 35.5 | 1.0 | 26 |
| TATG | 37.9 | 38.3 | 1.2 | 31 |
| TATT | 35.0 | 36.0 | 1.1 | 61 |
| TCAA | 43.7 | 43.0 | 1.1 | 30 |
| TCAC | 39.7 | 39.0 | 1.1 | 23 |
| TCAG | 46.3 | 44.8 | 1.3 | 35 |
| TCAT | 48.1 | 46.5 | 1.4 | 36 |
| TCCA | 43.7 | 42.6 | 1.3 | 28 |
| TCCC | 38.6 | 39.5 | 1.1 | 18 |
| TCCG | 39.3 | 40.8 | 1.6 | 32 |
| TCCT | 41.2 | 40.7 | 1.0 | 35 |
| TCGA | 46.4 | 45.2 | 2.0 | 42 |
| TCGC | 40.4 | 40.0 | 1.2 | 37 |
| TCGG | 40.0 | 41.0 | 1.5 | 30 |
| TCGT | 37.9 | 38.9 | 1.4 | 71 |
| TCTA | 42.1 | 41.3 | 1.2 | 33 |
| TCTC | 34.6 | 34.8 | 0.9 | 34 |
| TCTG | 32.8 | 33.1 | 1.3 | 23 |
| TCTT | 32.4 | 33.1 | 1.4 | 68 |
| TGAA | 45.6 | 46.0 | 2.1 | 31 |
| TGAC | 38.5 | 38.6 | 1.1 | 32 |
| TGAG | 42.4 | 43.1 | 1.2 | 24 |
| TGAT | 43.3 | 43.3 | 1.3 | 16 |
| TGCA | 42.8 | 42.4 | 1.7 | 28 |
| TGCC | 35.4 | 35.9 | 1.6 | 27 |
| TGCG | 40.6 | 40.4 | 1.7 | 43 |
| TGCT | 37.3 | 37.9 | 1.3 | 34 |
| TGGA | 38.7 | 38.6 | 1.3 | 27 |
| TGGC | 32.1 | 32.5 | 1.1 | 21 |
| TGGG | 33.7 | 33.8 | 1.0 | 14 |
| TGGT | 27.1 | 28.0 | 1.7 | 64 |
| TGTA | 27.1 | 27.0 | 1.6 | 27 |
| TGTC | 17.6 | 17.1 | 1.4 | 32 |
| TGTG | 21.7 | 21.7 | 1.1 | 19 |
| TGTT | 18.8 | 18.6 | 1.0 | 35 |
| TTAA | 40.1 | 40.4 | 1.6 | 39 |
| TTAC | 35.6 | 35.6 | 1.2 | 53 |
| TTAG | 41.8 | 41.2 | 1.3 | 53 |
| TTAT | 40.6 | 39.8 | 1.1 | 51 |
| TTCA | 36.0 | 34.8 | 1.6 | 40 |
| TTCC | 28.7 | 28.6 | 1.3 | 39 |
| TTCG | 33.2 | 32.5 | 1.6 | 62 |
| TTCT | 32.4 | 31.6 | 1.6 | 46 |
| TTGA | 33.1 | 32.8 | 1.2 | 30 |
| TTGC | 27.3 | 27.6 | 1.0 | 45 |
| TTGG | 32.1 | 32.0 | 1.0 | 46 |
| TTGT | 28.8 | 28.8 | 0.8 | 55 |
| TTTA | 29.2 | 28.3 | 1.3 | 67 |
| TTTC | 22.2 | 21.8 | 1.1 | 58 |
| TTTG | 25.9 | 26.2 | 0.9 | 63 |
| TTTT | 22.2 | 22.8 | 0.9 | 89 |

**Description of automatic level finding algorithm**
Our algorithm identifies levels in two steps. First, it identifies the level boundaries within the time-traces. Then it removes spurious levels and combines levels where the polymerase appears to have "toggled" between two bases.

**Identifying level locations.** Starting from the beginning of a current trace our technique first examines part of a current trace and divides it into two sections. Under the assumption that the sampled currents within each level are Gaussian-distributed, we compute the total probability that the two sections originated from two distinct Gaussian distributions. We divide this total probability by the probability of the null hypothesis, namely that the combination of the two sections originated from a single Gaussian. For ease of computation, we use log probabilities. For a given section the observed mean is $\hat{I}$ and width is $\sigma$. For an individual current measurement within this section, $I(t)$, at time $t$, the log probability (density) is:

$$\log p(I(t)) = \log\left(\frac{1}{\sqrt{2\pi}\sigma}e^{(I(t)-\hat{I})^2/2\sigma^2}\right) \qquad \text{Eq. 1.}$$

To find the total log probability that the observations between $t_1$ and $t_2$ belong to a single level defined by $\hat{I}$ and $\sigma$, we sum Eq. 1 between $t_1$ and $t_2$ and use the definition of sigma, giving

$$\log p(I([t_1, t_2])) = (t_2 - t_1)\log\sigma + \text{const} \qquad \text{Eq. 2.}$$

We calculate total log probability (Eq. 2) for each of the two sections between $t_1$ and $t_2$, and then between $t_2$ and $t_3$. To compare the log probabilities, we subtract the total log probability of the null hypothesis by computing Eq. 2 between $t_1$ and $t_3$. Combining all probabilities yields a comparison metric, which we denote log$p(t1,t2,t3)$,

$$\begin{aligned}\log p(t_1, t_2, t_3) =& (t_2 - t_1)\log\sigma(t_2, t_1) + (t_3 - t_2)\log\sigma(t_3, t_2) \\ & - (t_3 - t_1)\log\sigma(t_3, t_1) + \text{const}\end{aligned} \qquad \text{Eq. 3.}$$

The $t_2$ yielding minimal log$p(t_1, t_2, t_3)$ indicates the location of a possible level transition within the current observations between $t_1$ and $t_3$. In our level finding algorithm, we begin with a given time window ($[t_1, t_3]$) and search for $t_2$ that minimizes log$p$. If min(log$p$) is less than a specified threshold (we used a threshold of log$p$ = -50) there is a level transition at $t_2$, and we recursively search between $t_1$ and $t_2$ and between $t_2$ and $t_3$ for other level transitions. If min(log$p$) is above the threshold for the original time window then there are no transitions within $t_1$ and $t_3$ and we consider a larger window by increasing the value of $t_3$.

Not all levels correspond to the single nucleotide forward motion of the phi29 DNAP. We remove levels that (1) have durations < 500μs, (2) have mean values outside of expected ranges (10-70 pA) or (3) have an error in the mean greater than 5 pA. Finally, using the logp calculated using Eq. 3 above, we identify regions of levels that are similar, and combine the levels in them.

## Description of alignment algorithm

Our tool for aligning level sequences is based on the well-known Needleman-Wunsch and Smith-Waterman algorithms for sequence alignment (*21,22*). A Needleman-Wunsch or Smith-Waterman alignment of two base sequences A and B allows for gaps in both sequences. Due to possible gaps the optimal alignment between the first $n_A$ bases of A and the first $n_B$ bases of B is one of the following: 1) the optimal alignment between the first $n_{A-1}$ bases of A and all $n_B$ bases of B plus a gap in A; 2) the optimal alignment between all $n_A$ bases of A and the first $n_{B-1}$ bases of B plus a gap in B; or 3) the optimal alignment between the first $n_{A-1}$ bases of A and the first $n_{B-1}$ bases of B plus a final matched base in both sequences. Longer optimal alignments are recursively calculated from shorter optimal sub-alignments, and the entries in the alignment table (**sup. fig. 5a**) are filled from top-left to bottom-right.

The main complication in our experiment is that our level sequences sometimes step backwards. If we used the Needleman-Wunsch scheme directly each entry in the alignment table would depend on both the entry to the left (due to forward steps) and the entry to the right (due to backsteps), which in turn would depend on the entry in question. To fix this problem, we require each step in the alignment to advance sequence A forward by one, as shown in **Supplemental figure 5b**. Our alignment is therefore the optimal mapping of *every* level in sequence A to its corresponding level in sequence B (or a null level if no good match exists). Note that our alignment trace passes over rather than through skipped levels in sequence B (**Sup. fig. 5c**).

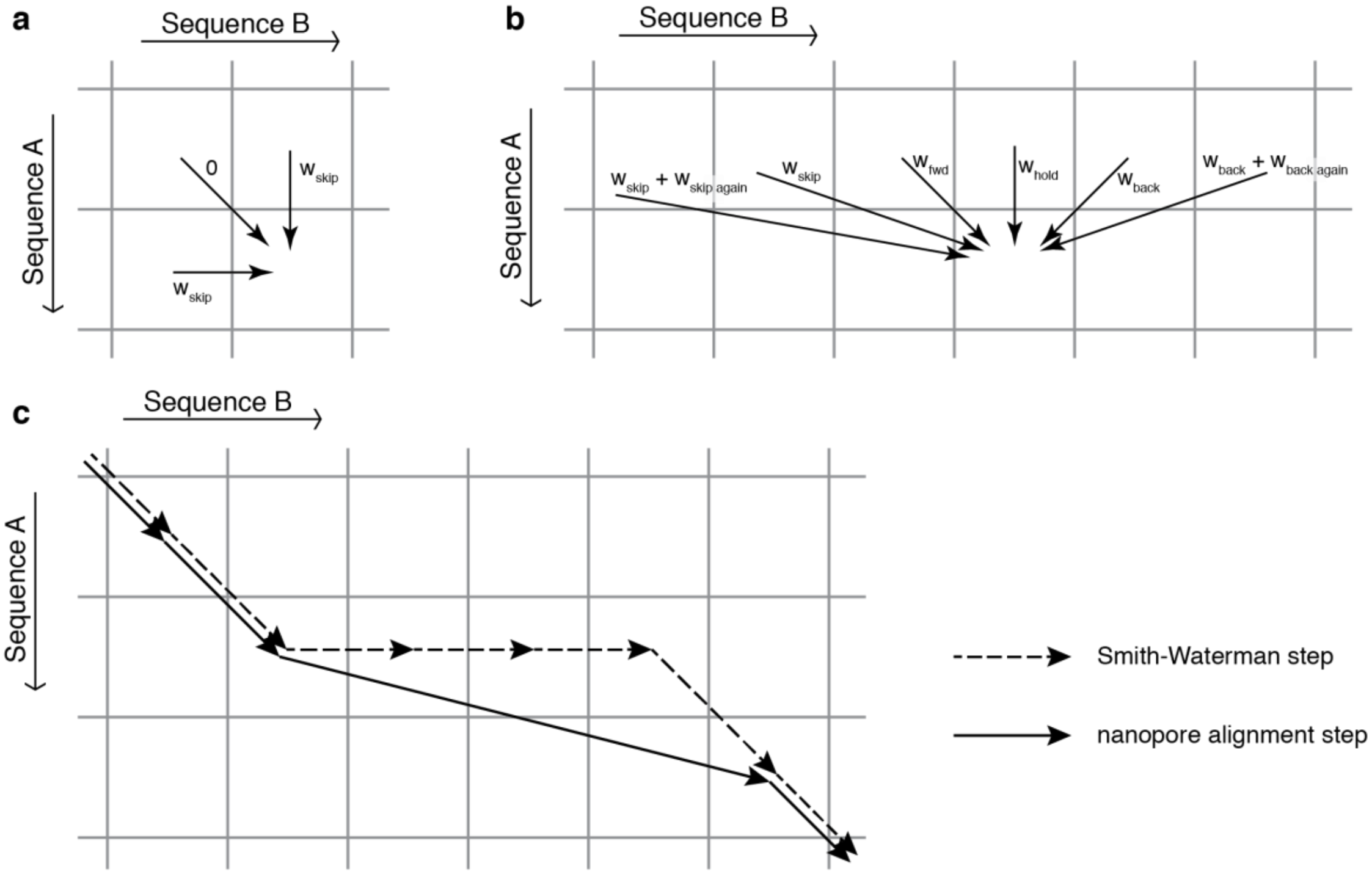

**Supplemental figure 5: a**) Needleman-Wunsch alignments consider horizontal and vertical steps in the alignment table (with a penalty w), corresponding to mismatched bases in one sequence or the other. Diagonal steps, indicating a matched base, generally have no penalty unless there is a mismatch. **b**) Our nanopore alignment forces every step to progress along sequence A, but allows backsteps in sequence B. We assign affine penalties to skips and backsteps: for example, a backstep of 3 levels would earn the backstep penalty $w_{back}$ plus twice the backstep-again penalty $w_{backagain}$. **c**) The difference between a Needleman-Wunsch alignment and our nanopore alignment when there is a skip in sequence B.

## Supplemental figure 6: Alignment of nanopore reads to quadromer prediction:

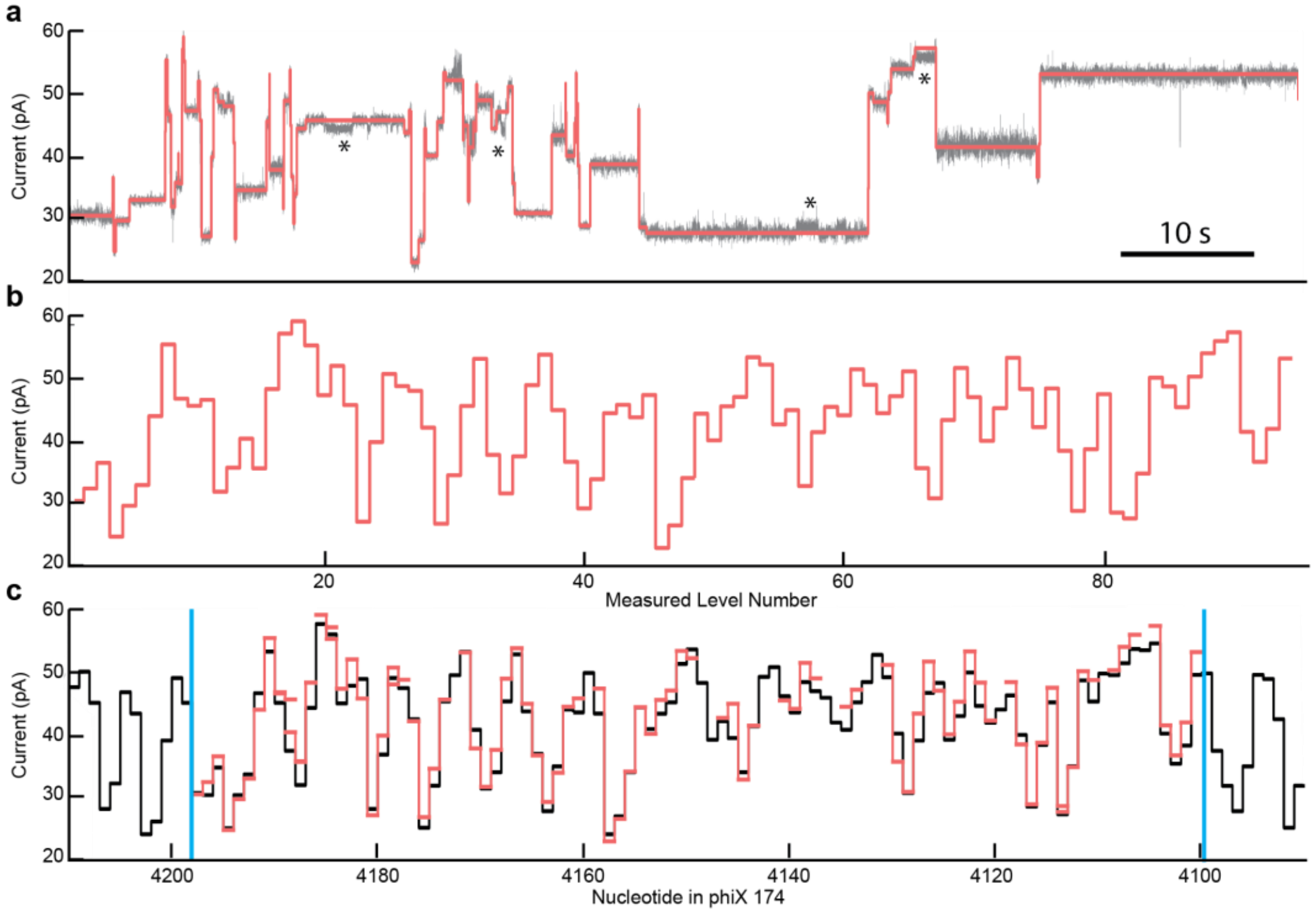

**Supplemental figure 6.** Similar to Fig. 3 in the main text, this figure demonstrates how alignment takes place. Instead of an amplicon being aligned, this figure shows a short sub-segment from a long event. **a**) Our level finding algorithm is used to identify transitions between levels in the current trace. Asterisks mark locations where the algorithm identifies and removes DNAP backsteps. **b**) We then extract the sequence of median current values from each level. **c**) Next, we align the current values to predicted values from the reference sequence using a dynamic programming alignment algorithm similar to Needleman-Wunch alignment (*21*). In some locations, levels are skipped in the nanopore read either due to motions of the DNAP or errors by the level finding algorithm. We determine read boundaries from the first and last matched levels in the reference sequence. Read boundaries are indicated by the blue lines.

## Calculating alignment significance

Using the afore mentioned alignment algorithm we align nanopore reads to the predicted level sequence from a known DNA sequence using the quadromer map. The alignment produces a raw score, s, that can be compared to alignments to other reference sequences. Next, we generate a large random sequence along with the expected current levels. We then perform alignments of our measured data to the random sequence. If our measured levels are truly from phi X 174 we expect the score to stand out from the distribution of scores to random alignments. **Supplemental figure 7** shows a histogram of the scores for random alignments (blue) and a marker (red) for the location of the score for the alignment to phi X 174. Strongly negative scores represent good alignments.

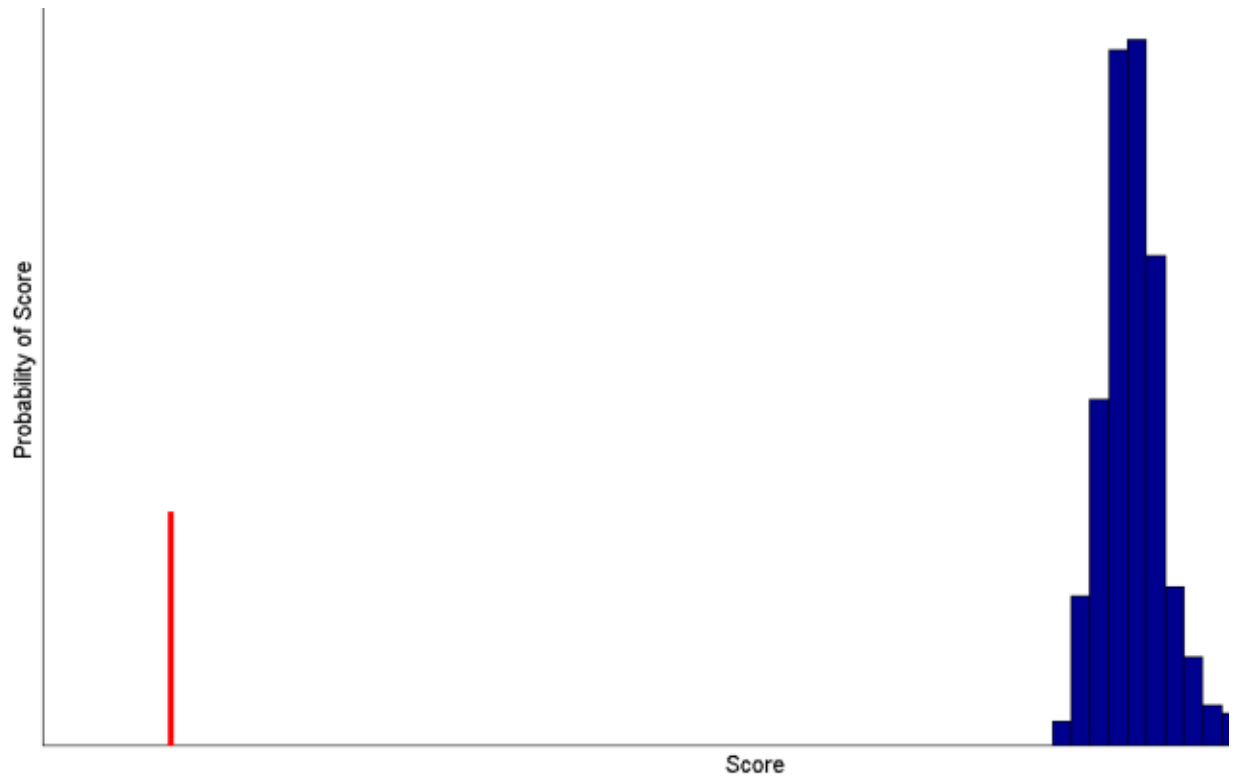

**Supplemental figure 7**: The probability distribution of scores for random alignments dP/dS (blue), together with a marker (red) showing the location of the score of the alignment to phi X 174. Units are in arbs. This plot is made using event number 49 from figure 4 of the main

The confidence in the alignment $C$ is calculated by

$$C(S_0) = \int_{-\infty}^{S_0} \frac{dP}{dS} dS$$
.

$C$ represents the probability that a given alignment to a random sequence will produce a score better than $S_0$. This particular alignment had a confidence score of $10^{-147}$, reflecting a high probability of these measured levels belonging to phi X 174 relative to random alignments. We assume that in the limit of an infinite number of random alignments, the distribution of alignment scores for random sequences approximate a Gaussian, so that

$$C(S_0) = \int_{-\infty}^{S_0} G(S) dS$$
.

We find the Gaussian by fitting to the width and center of the measured score distribution as floating parameters.

We comment briefly on the meaning of $C$ because of the extreme smallness of these numbers. $C$ represents nothing more than the probability that the produced alignment could also be randomly obtained. The score of $10^{-147}$ was produced by an alignment of nanopore read of length ~2000.

**Supplemental figure 8: Coverage plot for phi X 174 amplicons**

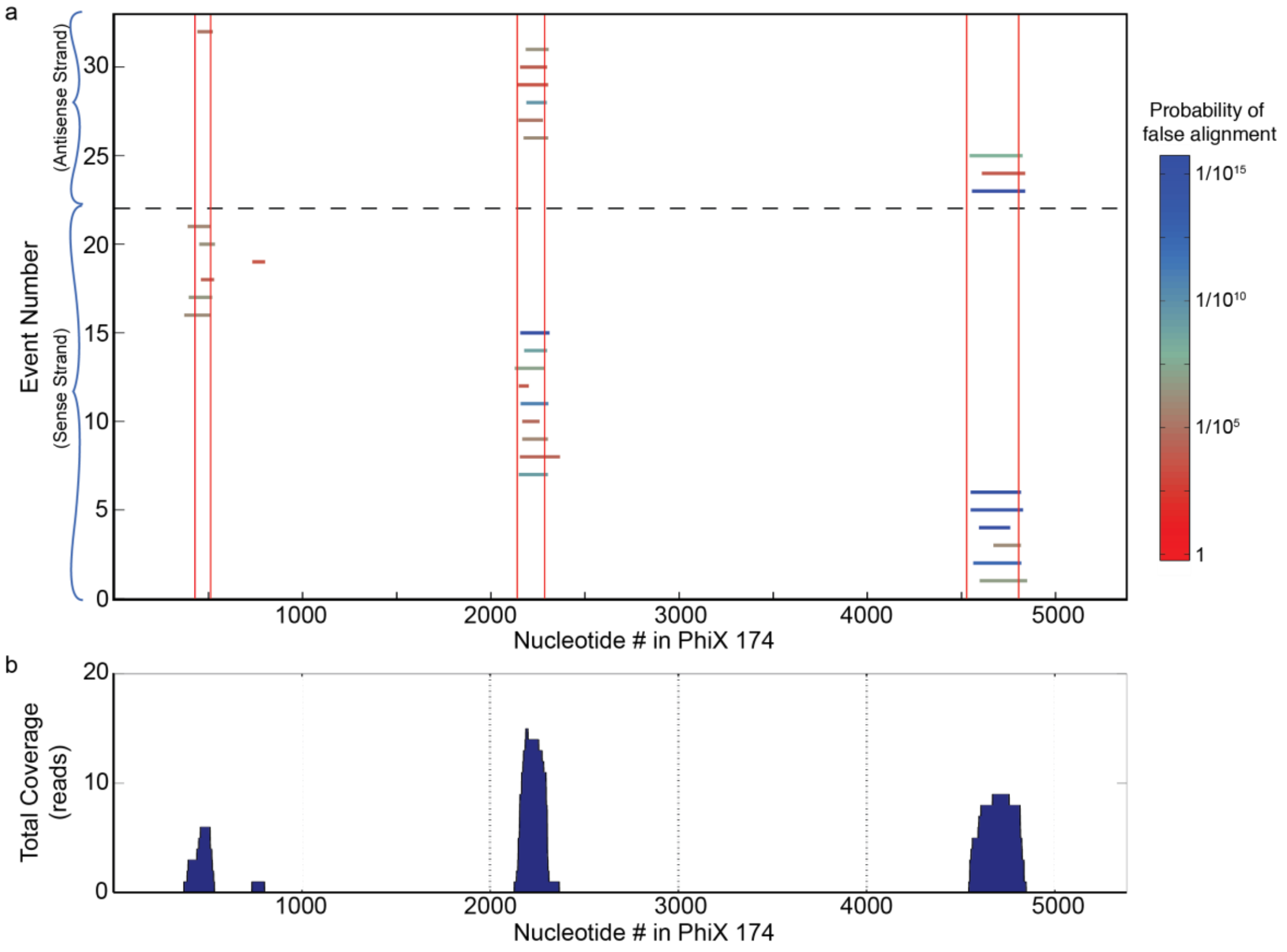

**Supplemental figure 8: Amplicon alignment.** 31 reads of phi X 174 amplicons aligned to current reference generated by translating the known phi X 174 sequence into current levels using the quadromer map. DNA strands are identified with high confidence, which enables a number of different useful applications such as organism identification and providing a reconstruction scaffold for short high-quality reads obtained with other sequencing technologies. **a**) Alignment bounds for 31 nanopore reads of phi X 174 amplicons. The alignment bounds match well with the actual amplicon locations. All reads with a quality better than 1 in $10^4$ fall within one of three locations along the phi X 174 genome revealing the correct location of the amplicons within the genome. Because the adaptors attach to the strands in random orientation, we made reads of both the sense and anti-sense strands. **b**) Coverage for nanopore reads in **a**).

**Full phi X 174 library gel**

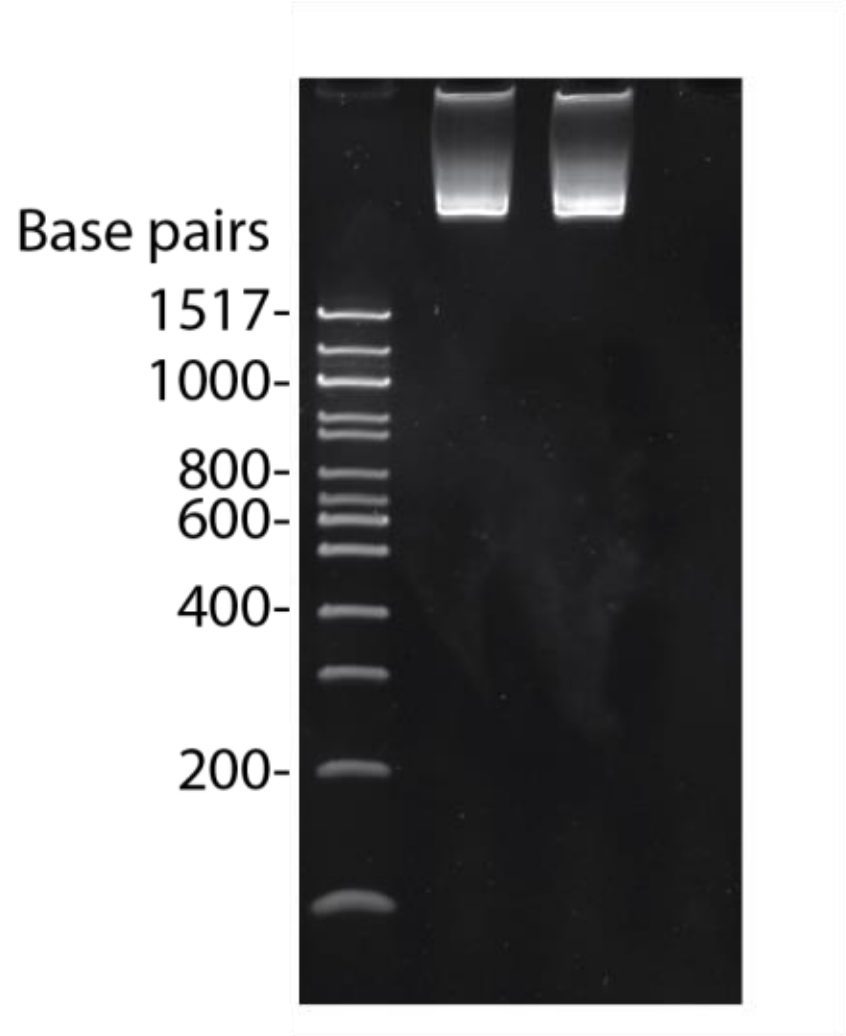

**Supplemental figure 9:** The gel shows the length distribution of phi X 174 genomic DNA used in our experiments (lanes 2 & 3 are replicates). There is a single band that contains broadening towards longer strand lengths. The band broadening is likely due to different numbers of ligated adaptors (0,1, or 2), different adaptor orientations (fan-tail or hairpin), or possibly circular forms of phi X 174. The absence of bands of shorter lengths indicates that our nanopore read lengths are determined not by library quality, but instead by how far the enzyme processes along the DNA before dissociating.

## Phi X 174 consensus and quadromer map revision

We used values from a current level consensus for the phi X 174 genome to update the quadromer map. To generate the consensus current level sequence for phi X 174, we aligned each nanopore read of the phi X 174 DNA to the predicted current level sequences for its sense and antisense base sequences. The predicted current level sequences were made from the initial measurements of quadromers in the de Bruijn sequence. Alignments with an overall confidence better than $10^{-6}$ were selected to contribute to the updated map.

Only levels aligned with high certainty contributed to the consensus. All consensus level values were the average of at least four reads. Also, at least half of all reads covering this level contained a current level matched to that predicted level. The consensus value for the given context was calculated as the mean of all reads aligned to that level.

With the exception of the self complimentary quadromer GATC, there are many instances (35 on average, see **sup. fig. 10**) of each of the remaining 255 quadromers within the 5386-nucleotide phi X 174 genome and its complementary strand. For GATC, we retain the original de Bruijn sequence current value. Using the updated consensus levels, we were able to update the quadromer map with additional measurements in a variety of sequence contexts. The revised map uses the mean and standard deviation of all measurements made of each quadromer throughout the phi X 174 consensus (Table S2). **Supplemental figure 11** (next page) shows the revised quadromer map in comparison to the original quadromer map.

The consensus generation and quadromer map updating procedures were tested by reserving five high-quality reads to be excluded from the generation of the revised quadromer map, and then aligning these reads to both the consensus sequence and the updated prediction. In all cases, the confidence in the alignments of these reads improved dramatically when the new prediction was used as the reference sequence.

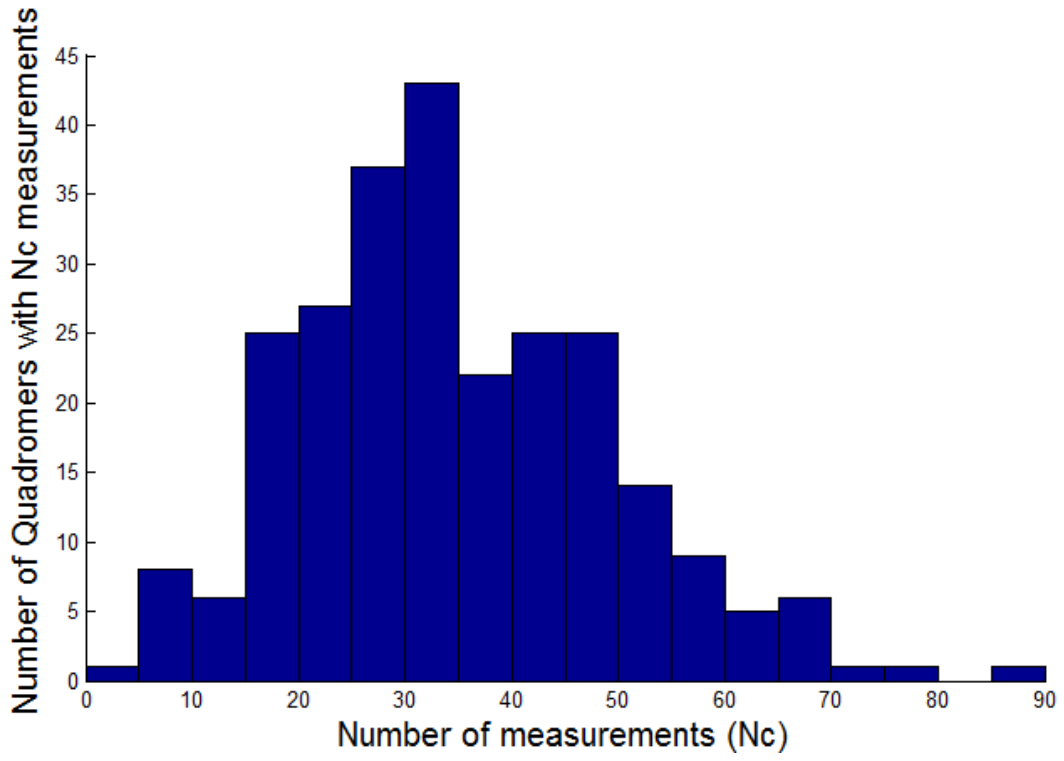

**Supplemental figure 10**: Histogram of the number of instances of each quadromer in the phi X genome. Each quadromer has 35 reads on average.

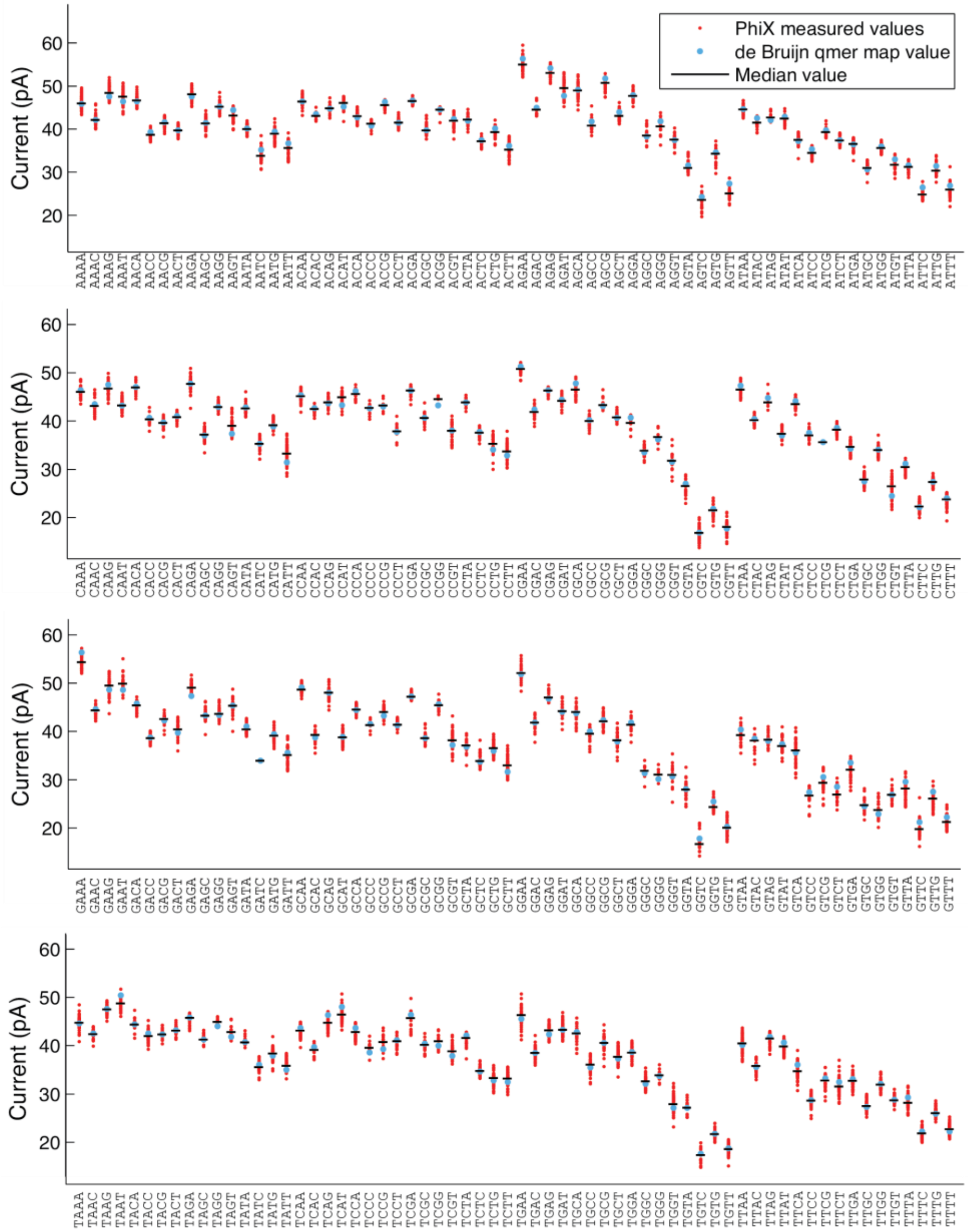

**Supplemental figure 11:** The revised quadromer map in alphabetical order, with 64 quadromers (written 5'-3') in each panel, beginning with A,C,G,T, respectively. Red dots are measured values of the quadromer in phi X 174. The blue dots are the values from the original de Bruijn quadromer map. The black lines are the medians of all measured instances of a given quadromer.

## DNA scaffold reconstruction:

The difficulty of *de novo* sequencing with most sequencing technologies is that their many short DNA reads must be stitched together in the proper order to form a long contiguous sequence. This assembly process is usually performed by looking for sequence similarity between overlapping reads. We demonstrate an alternative method of sequence scaffolding by mapping 100 short, 100 bp long reads from an Illumina MiSeq sequencer to one of our long (3466 levels) nanopore read. The mapping was performed by converting the sequence of each Illumina read, and its reverse compliment, into a sequence of current levels using our quadromer map, and then using our level alignment tool to find the likely location of the current level sequence in our nanopore read. Figure S13 shows the fate of the 87 (out of 100) Illumina reads which generated an alignment to the nanopore read: 61 Illumina reads lay at least partially within the nanopore read and were aligned properly; 9 Illumina reads lay at least partially within the nanopore read and were misaligned; and 17 Illumina reads fell entirely outside the nanopore read. 9 of the 13 Illumina reads that did not generate an alignment actually lay outside the nanopore read.

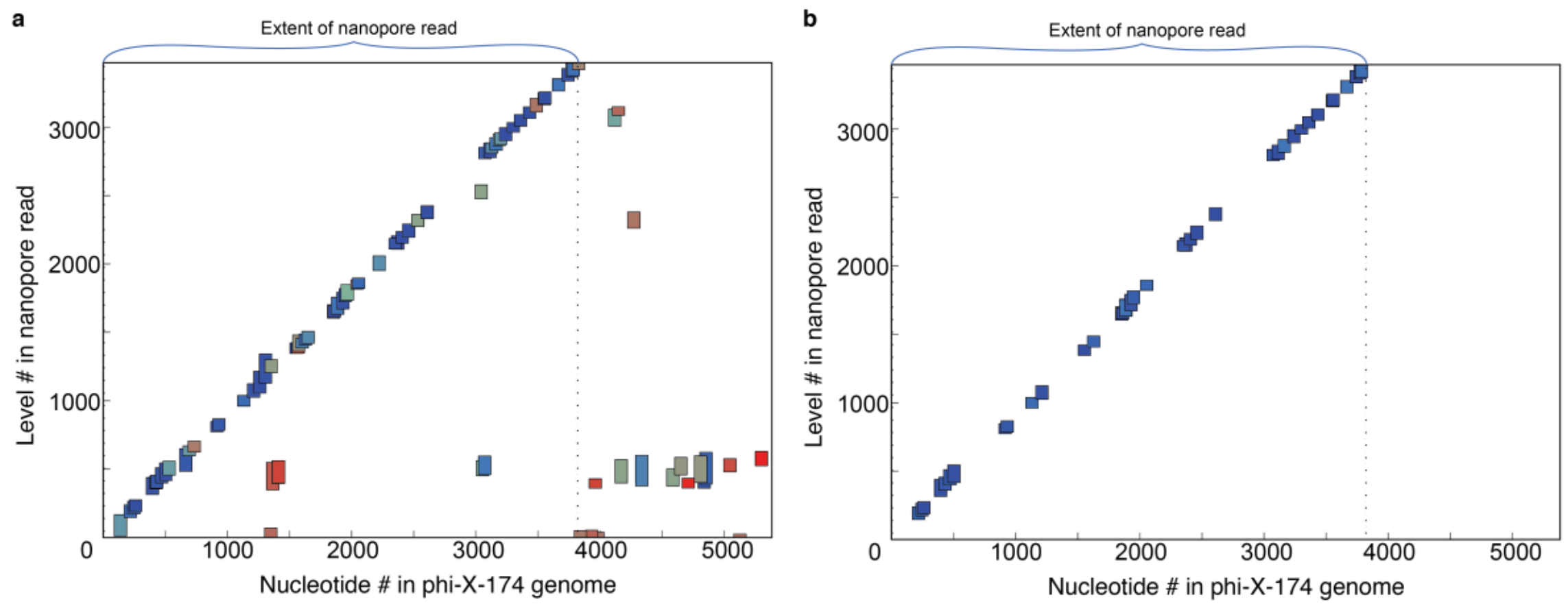

**Supplemental figure 12**: Alignment-scaffolded assembly of 87 short DNA sequences. Each short DNA sequence is indicated by a box, whose horizontal width indicates the location of the Illumina read within the phi X 174 genome and whose vertical height indicates the span of the Illumina read alignment to a 3466 level nanopore read (spanning 3819 bp in the phi X 174 genome). **a**) Location of all 87 reads that produced alignments to the nanopore read. Color indicates the alignment quality: blue is high-quality and red is low-quality. Overlapping rectangles represent contigs. **b**) After applying a cutoff filter on nanopore alignment quality and the length of the alignment to the nanopore read (keeping only alignments spanning less than 130 levels) we see that all erroneous alignments are filtered out (plus 23 low scoring but correct alignments). Of the 74 Illumina reads which should have aligned to our nanopore read, we are left with 38 (51%) Illumina reads properly localized within the phi X 174 genome with high confidence.

**Viral alignment and identification:**

Viral identification was performed by aligning arbitrary 250 level subsets of nanopore reads of phi X 174 DNA to a viral database consisting of 5287 viruses (including phi X 174) totaling 156 megabases. To increase the alignment speed, first an 80-level subset of the nanopore read was aligned to every viral genome in the database. This initial alignment was used to generate a list of likely candidate alignments. Alignment confidences for each 80-level alignment to each virus were tallied and compared. Viruses with log confidences better than the mean log confidence score by 3 standard deviations were passed on to the next round; all others were discarded. In the next alignment round, 150 levels of the nanopore read were aligned to the remaining viruses followed by another round of database reduction. Finally all 250 levels were aligned to the remaining viruses. For each event tested, the 250-level alignment correctly identified the DNA as belonging to phi X 174 with at least 99.9996% confidence (in all instances, the 80 or 150-level alignment also suggested phi X 174 as the most likely although with reduced confidence).

Performing a final alignment of the entire >1000-level nanopore reads to the phi X 174 genome can confirm the conclusion to almost arbitrarily high confidence (less than 1 in $10^{70}$ chance of mis-identification).

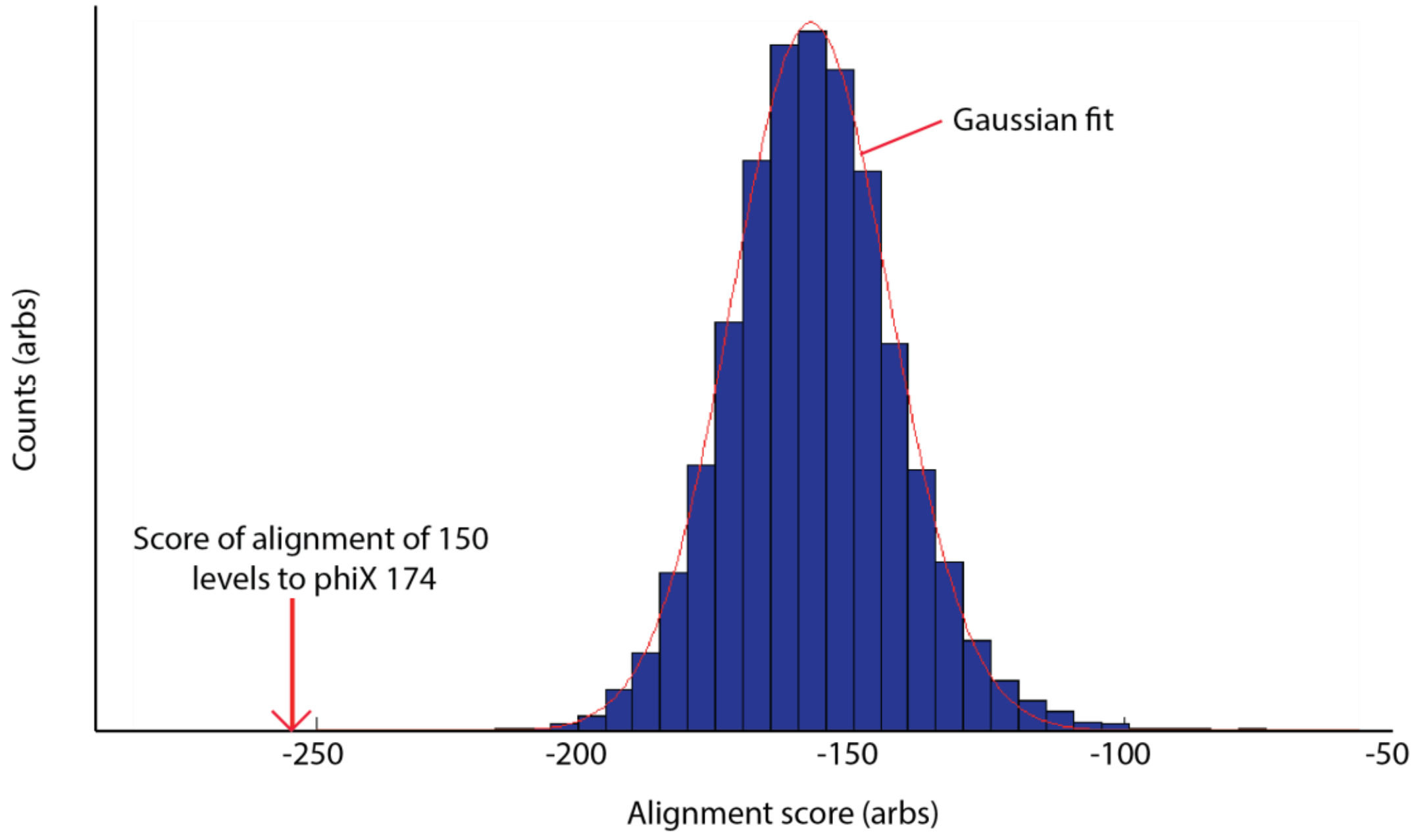

**Supplemental figure 13:** Distribution of alignment scores for a 150-level segment from a long nanopore read to a viral genome database. The distribution of scores is Gaussian. Here the 150-level alignment to phi X 174 differs from random alignments by ~6.5 standard deviations, identifying the strand with 99.9999997% confidence.

## SNP calling workflow schematic

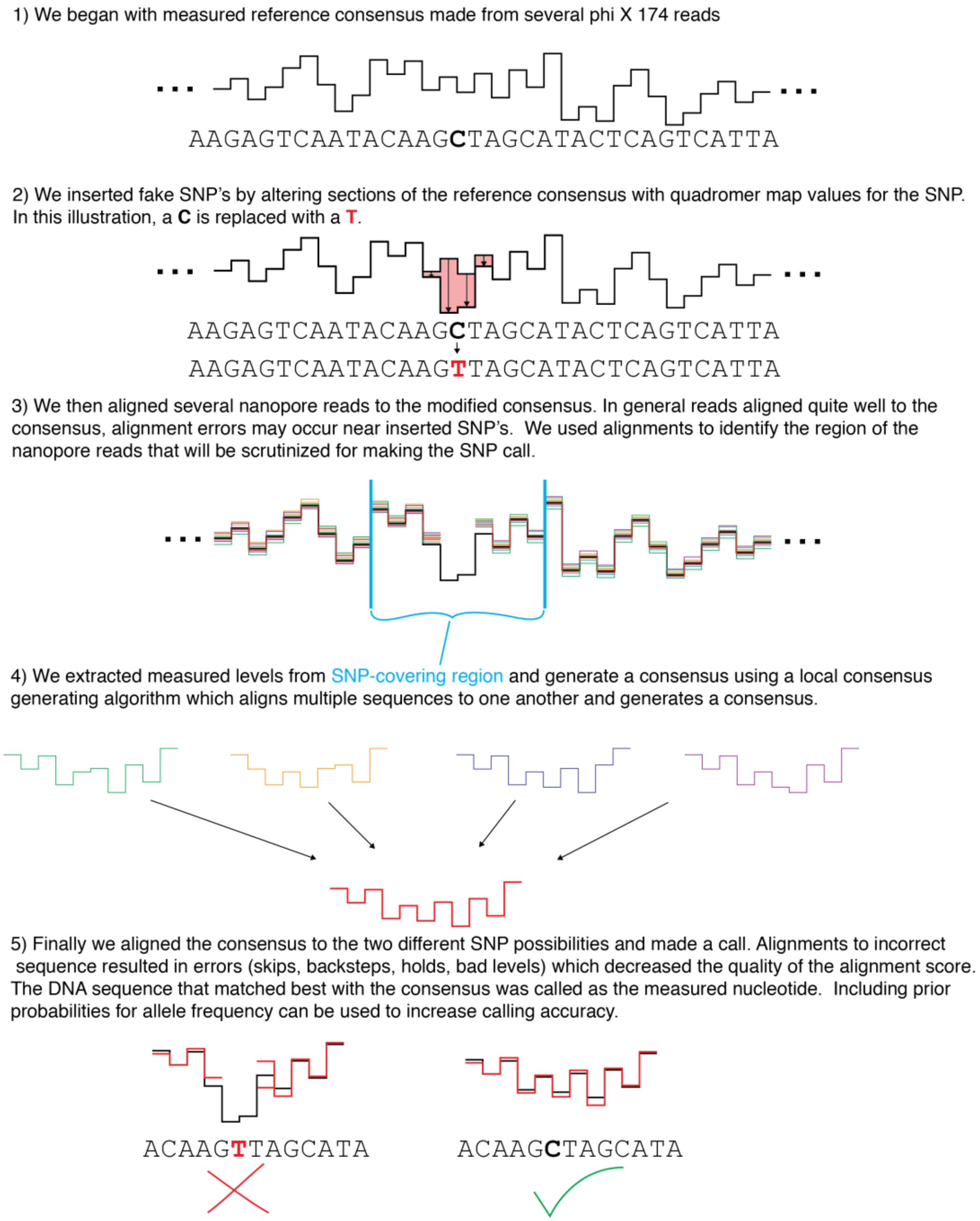

**Supplemental figure 14**: Schematic outline of SNP detection methods. We inserted “mock SNPs” into a reference consensus by inserting quadromer map values corresponding to the inserted SNP. Transversions and transitions were inserted into the genome in the following ratio (70% C<–>T/G<–>A, 15% C<–>A/G<–>T, 9% G<–>C, and 5% A<–>T) corresponding to how often they occur within the human genome (34). We then performed alignments of nanopore reads to the reference consensus as if we were comparing new nanopore reads to a previously measured consensus. We used these alignments to extract current levels from events that had reads of the SNP region in question. We then generate a consensus using these nanopore measurements. The sequence that aligns best with the consensus is selected as the measured allele. See **Sup. Fig. 16** for detection efficiencies.

## SNP detection efficiencies and resequencing confusion matrix

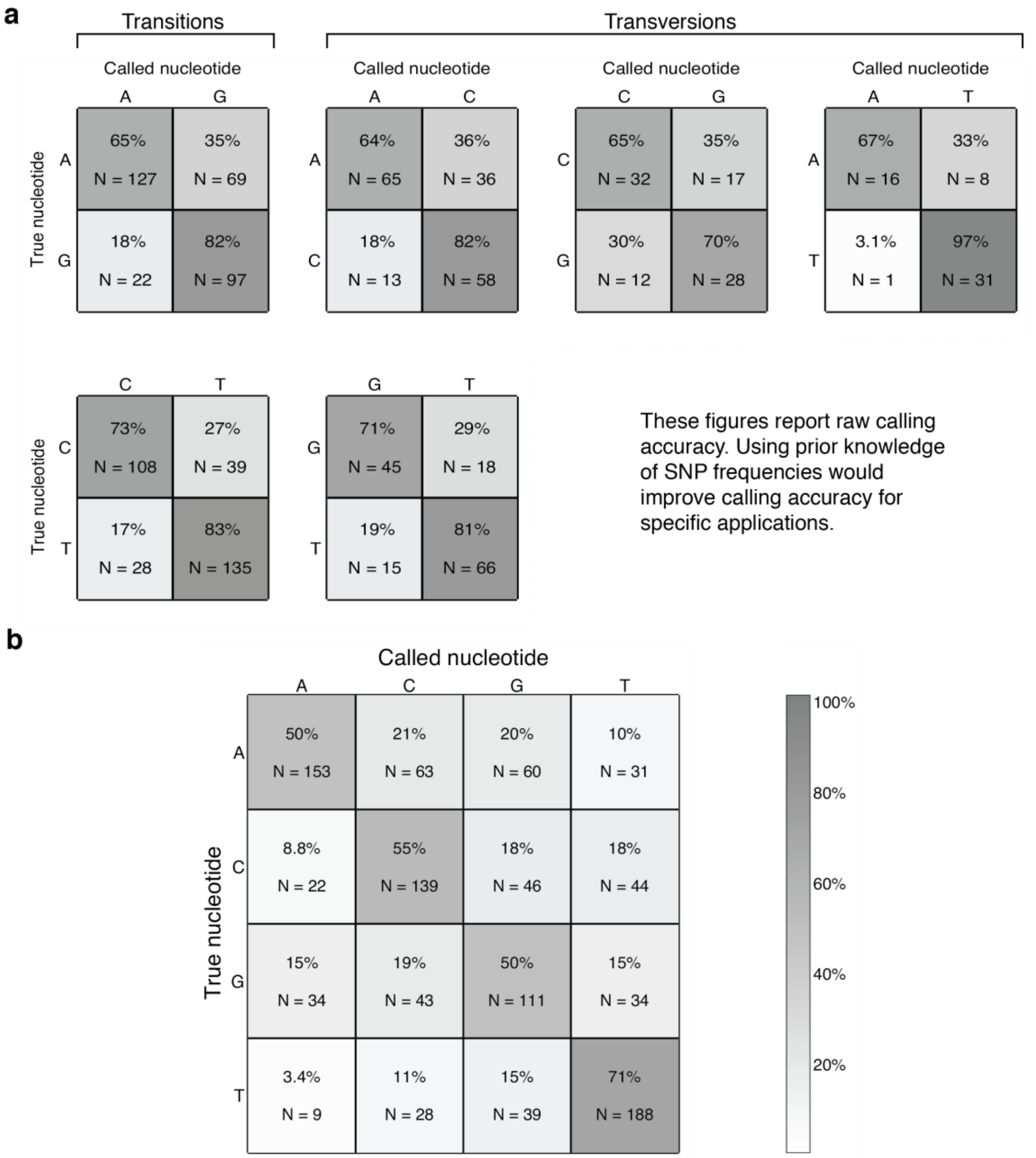

**Supplemental figure 15: Confusion matricies for SNPs and reference sequencing. a**) Detection efficiencies for each possible SNP in each box. The actual DNA nucleotide is displayed along the left of each box while the nanopore call is displayed along the top of each box. The contrast within the box indicates our ability to distinguish between the two nucleotides in various sequence contexts. **b**) shows detection efficiency for reference sequencing where instead of comparing only two nucleotides (as one does when interrogating most SNPs), we select the nucleotide that matches the data best out of all four nucleotides. Combining reads of both sense and anti-sense strands can increase calling accuracy.